\documentclass[sigconf,anonymous=false,review=false,screen=true]{acmart}

\usepackage[utf8]{inputenc}
\usepackage{textcomp}
\usepackage[american]{babel}
\usepackage{xspace}
\usepackage{balance}
\usepackage{booktabs} 
\usepackage[T1]{fontenc}
\usepackage{lmodern}
\usepackage{tabularx} 
\usepackage{footnote}
\usepackage{ltablex}
\usepackage{booktabs}
\usepackage{caption}
\usepackage{subfig}
\usepackage{rotating}
\usepackage{framed}
\usepackage{color}
\usepackage{graphicx}
\usepackage{makecell}
\usepackage{listings}
\usepackage{xfrac}
\usepackage{wrapfig}
\usepackage{caption}
\usepackage{soul}
\usepackage{etoolbox}
\usepackage{float}
\usepackage{colortbl}

\definecolor{Gray}{gray}{0.85}

\renewcommand\footnotetextcopyrightpermission[1]{} 
\begin{document}
\title[Increasing  the Quality of 360° Video Streaming]{Increasing  the Quality of 360° Video Streaming by Transitioning between Viewport Quality Adaptation Mechanisms}
\author{
	\hspace{10pt} Christian Koch\textsuperscript{1}, Arne-Tobias Rak\textsuperscript{1}, Michael Zink\textsuperscript{2}, Ralf Steinmetz\textsuperscript{1}, Amr Rizk\textsuperscript{3}}

\affiliation{
	\institution{\vspace{0pt}
		\textsuperscript{1}Multimedia Communications Lab (KOM), Technische Universit\"at Darmstadt, Germany\\
		\textsuperscript{2}University of Massachusetts Amherst, USA, 
		\textsuperscript{3}Ulm University, Germany
	}
}
\email{christian.n.koch@gmail.com, arne-tobias.rak@stud.tu-darmstadt.de, zink@ecs.umass.edu}
\email{ralf.steinmetz@kom.tu-darmstadt.de, amr.rizk@uni-ulm.de}

\begin{abstract}
Virtual reality has been gaining popularity in recent years caused by the proliferation of affordable consumer-grade devices such as Oculus Rift, HTC Vive, and Samsung VR.
Amongst   the various VR applications, 360° video streaming is currently one of the most popular ones.
It allows user to change their field-of-view (FoV) based on head movement, which enables them to freely select an area anywhere from the sphere the video is (virtually) projected to.
While 360° video streaming offers new exciting ways of consuming content for viewers, it poses a series of challenges to the systems that are responsible for the distribution of such content from the origin to the viewer.
One challenge is the significantly increased bandwidth requirement for streaming such content in real time.
Recent research has shown that only streaming the content that is in the user’s FoV in high quality can lead to strong bandwidth savings.
This can be achieved by analyzing the viewers head orientation and movement based on sensor information.
Alternatively, historic information from users that watched the content in the past can be taken into account to prefetch 360° video data in high quality assuming the viewer will direct the FoV to these areas.
In this paper, we present a 360° video streaming system that transitions between sensor- and content-based predictive mechanisms.
We evaluate the effects of this transition-based approach on the Quality of Experience (QoE) of such a VR streaming system and show that the perceived quality can be increased between 50\% and 80\% compared to systems that only apply either one of the two approaches.
\end{abstract}

%
%

\ccsdesc[500]{Information systems~Multimedia streaming}
\ccsdesc[500]{Computing methodologies~Virtual reality}
\ccsdesc[100]{Computing methodologies~Video segmentation}

\keywords{360° Video Streaming, Virtual Reality, Quality Adaptation}

\maketitle

\section{Introduction}\label{sec:intro}

Large events such as sports matches or the keynotes of Apple and Google are nowadays regularly captured in 360°.
These videos are created by stitching multiple camera perspectives together and projecting the result on a sphere with the viewer at its center~\cite{el2016streaming}.
Such videos can be watched by panning in a traditional web browser's video player or by using VR goggles.
Especially the latter ones create an immersive experience since the user's head movement directly affects the visible area of the video, i.e., the field-of-view (FoV).

The proliferation of VR devices such as Oculus Rift\footnote{\url{https://www.oculus.com/rift/} [Accessed: \today]}, HTC Vive\footnote{\url{https://www.vive.com/} [Accessed: \today]}, and Samsung VR\footnote{\url{https://www.samsung.com/us/mobile/virtual-reality/} [Accessed: \today]}  makes 360° video streaming available to a broad audience since they are affordable also for individual users.
In addition, 360° content is offered on YouTube and other streaming websites.
To ensure a reasonable Quality of Experience (QoE) for viewers of 360° videos, a high resolution and frame rate is required which also helps preventing motion sickness \cite{corbillon2017viewport}.
Transferring the entire 360° video in high resolution is particularly bandwidth-consuming.
360° videos reach up to 12k, i.e., 1520x6480 pixel in resolution with 100 frames per second (FPS)~\cite{corbillon2017viewport}.
In many countries the average Internet speed into households is not sufficient for this task, e.g., in Germany it is just 15.3 Mbit/s~\cite{akamai}.
This is just a fraction of Google's recommended bitrate of 68\,Mbit/s for  for 4k 60 FPS 360° videos.\footnote{\url{https://goo.gl/gz556T} [Accessed: \today]}
%
%
Therefore, today the average household is incapable of streaming the entire sphere of a 360° video at high quality due to a lack of bandwidth, and streaming several 360° videos in parallel to different household member is impossible.
Optimizing bandwidth utilization for 360° video streaming is a recent research topic \cite{bao2016viewing,graf2017towards,qian2016optimizing,nasrabadi2017adaptive}.
%
One way to lower the bandwidth demand is to stream just the portion of the video that is currently visible to the viewer in a high resolution and the remainder in a low resolution~\cite{corbillon2017viewport}.
This currently watched portion is called viewport and may constitute about 20\% of the entire 360° sphere~\cite{bao2016viewing}.
However, this approach has two drawbacks:
i) A quick response time of about 10\,ms to head movements is crucial to maintain a high QoE~\cite{chakareski2018viewport}.
Usually, the Round Trip Time (RTT) between client and server is too high to allow for a sufficiently fast quality adaptation.
ii) Extracting the visible area from a video is a computationally expensive task that is hard to support by streaming servers~\cite{corbillon2017viewport}.
%
%
%
Multiple viewport-oriented schemes have been proposed to cope with these issues~\cite{nguyen2017new, qian2016optimizing, bao2016viewing, xie2017360probdash}.
One approach is to predict the user's head movement in order to request the new viewport in a timely fashion. 
Although this prevents rebuffering to some extent, it is only reliable for the duration for which the viewport can be accurately predicted
which is claimed to be up to
2\,s~\cite{qian2016optimizing}.
Consequently, maintaining useful content in the client's playout buffer in 360° streaming systems is difficult in case of fluctuating bandwidth.
Furthermore, abrupt changes in head position cannot be predicted which results in playback stalling.

To reduce computational stress on the server's side, the majority of the proposed schemes suggest to encode 360° videos into spatial and temporal segments such that segments corresponding to the current viewport can be requested instead of extracting them from the original source in an online manner.
Albeit not saving as much bandwidth as the previous approach, a more robust way of streaming 360° videos is to have the entire area around the viewport transferred in a lower quality than the viewport itself.
This enables buffering more content, since an incorrectly predicted viewport causes only a low-quality video segment to be displayed, instead of stopping the playback until the new segment is downloaded.
Several proposals have been published that incorporate transferring entire frames with a high quality at the viewport and its proximity~\cite{petrangeli2017improving, nasrabadi2017adaptive, chakareski2018viewport, corbillon2017viewport}.
Although helpful for maintaining a high QoE, viewport prediction is not mandatory in these streaming schemes and not used in all cases~\cite{facebook2016next}.
Some proposals introduce additional \emph{content-based prediction} mechanisms to further assist and improve accuracy.
This includes learning head movements from user traces for individual videos as well as identifying dynamic areas of a video which a viewer is likely to look at, i.e., saliency maps \cite{fan2017fixation}.

In this paper, we propose a novel approach for 360° video streaming system which uses, both, sensory and content-based predictive mechanisms.
We propose transitioning between these two mechanisms based on the client-server connection quality, similar to the self-adaptive concept from~\cite{frommgen2015aims}.
Additionally, an intermediate network cache is integrated to assist the transmission of video data.
We assume this cache to be a telco-owned edge cache that can be rented, e.g., a dedicated cache storage space per video.
Thus, the contributions of this paper are the design of a transition-based approach to 360° video streaming, as well as an evaluation of the effects of this scheme on video quality, bandwidth savings, and resulting QoE.

The remainder of  this paper is structured as follows.
Section~\ref{sec:relwork} discusses the related work on adaptive 360° video streaming.
In Section~\ref{existing-software}, we briefly summarize the head movement dataset and its software which we use in this paper.
We present our system design in Section~\ref{sec:system-design} and show our evaluation results in Section~\ref{sec:eval}.
Section~\ref{sec:discussion} discusses the proposed system.
We conclude the paper and provide an outlook on future work in Section~\ref{sec:conclusion}.

\begin{table*}
	\caption{Overview of existing works in the area of 360° video streaming.}
	\label{tab:relwork}
	\small
	\begin{tabular}{llllllllllll}
		& & \cellcolor{Gray} & \cellcolor{Gray}\hspace{-0.7cm}Adaptation based on & \cellcolor{Gray} & & & & & & & \\
		\rotatebox{0}{Reference} & \rotatebox{45}{VP-limit \hspace{-1cm}} & \rotatebox{45}{Sensor pred.} & \rotatebox{45}{Content pop. \hspace{-.5cm}} & \rotatebox{45}{Avail. Bandwidth \hspace{-2.2cm}} & \rotatebox{45}{Transitions \hspace{-1.2cm}} & \rotatebox{45}{Tiled/Monolithic \hspace{-.6cm}} & \rotatebox{45}{Caching} & \rotatebox{45}{Encoded Offline \hspace{-2cm}} & \rotatebox{45}{Codec} & \rotatebox{45}{Projection} & \rotatebox{45}{HTTP} \\ \hline \hline
		Petrangeli et al.~\cite{petrangeli2017http} 		& no & velocity & - & yes & no & tiled & - & yes & HEVC & equirect & 2 \\ \hline
		Nasrabadi et al.~\cite{nasrabadi2017adaptive} 	& no & \makecell[cl]{linear \\regression} & - & no & no & tiled & base layer & yes & SHVC & cubemap & 1.1 \\ \hline
		Fan et al.~\cite{fan2017fixation} 			& no & \makecell[cl]{neural\\ network} & \makecell[cl]{saliency- \& \\motion-map} & no & no & tiled & - & yes & HEVC & equirect & - \\ \hline
		Corbillon et al.~\cite{corbillon2017viewport} 	& no & - & - & yes & no & monolithic & - & yes & HEVC & arbitrary & 1.1 \\ \hline
		Kuzyakov et al.~\cite{facebook2016next} 		& no & - & - & yes & no & monolithic & - & yes & - & pyramid & 1.1 \\ \hline
		Chakareski et al.~\cite{chakareski2018viewport} 	& no & - & heatmap & yes & no & tiled & - & yes & HEVC & equirect & - \\ \hline
		Xu et al.~\cite{xu2018probabilistic} 		& no & \makecell[cl]{linear \\regression} & - & no & no & monolithic & - & yes & H.264 & pyramid & 1.1 \\ \hline
		Xie et al.~\cite{xie2017360probdash} 		& yes & \makecell[cl]{linear \\regression} & - & yes & no & tiled & - & yes & H.264 & equirect & 1.1 \\ \hline
		Nguyen et al.~\cite{nguyen2017new} 			& yes & velocity & - & yes & no & tiled & - & no & HEVC & equirect & - \\ \hline
		Qian et al.~\cite{qian2016optimizing} 		& yes & \makecell[cl]{linear \\regression} & heatmap & no & no & tiled & - & yes & H.264 & equirect & 1.1 \\ \hline
		Bao et al.~\cite{bao2016viewing} 			& yes & \makecell[cl]{neural\\ network} & - & no & no & - & - & no & - & equirect & - \\ \hline
		\rowcolor{Gray} \emph{our approach} & no & \cellcolor{Gray}{\makecell[cl]{linear \\regression}} & heatmap & \footnotemark & yes & tiled & \makecell[cl]{popular tiles} & yes & H.264 & Equirect & 1.1 \\
	\end{tabular}
\end{table*}

\section{Adaptive 360° Video Streaming}\label{sec:relwork}
A variety of approaches for 360° video streaming has been proposed.
First, we introduce the key properties of 360° video streaming systems to guide the classification of the related work.
We provide a comparative overview of selected and relevant related work in Table~\ref{tab:relwork}.

\subsection{Viewport-limited Approaches}
Viewport-limited approaches limit the transmitted video area to what is expected to be visible to the viewer.
This usually saves more bandwidth than transferring frames capturing the entire 360° sphere with emphasized quality regions, since less data needs to be transferred.
As viewports cover only about 20\% of a sphere, the remaining 80\% can theoretically be omitted~\cite{bao2016viewing}.
However, an accurate viewport prediction is mandatory in order to request the correct video areas from the server.
Incorrectly predicted viewports will either cause rebuffering or result in displaying a blank area to the viewer which significantly decreases the user's QoE~\cite{taghavinasrabadi2017adaptive}.
To reduce the risk of stalling, viewport-limited streaming schemes request additional video data, often in lower quality, in the proximity of predicted viewports, depending on the uncertainty of the prediction.
Furthermore, videos can only be buffered for the duration for which a sufficiently accurate prediction is possible. According to~\cite{qian2016optimizing},  a between 0.5\,s - 2\,s is sufficient.

\subsection{Adaptation Trigger}
Next, we present adaptive approaches that regulate the video quality using either the bottleneck bandwidth measured by the client, the user's head movements, or content popularity.

\begin{figure}[b]
	\centering
	\includegraphics[width=0.9\linewidth]{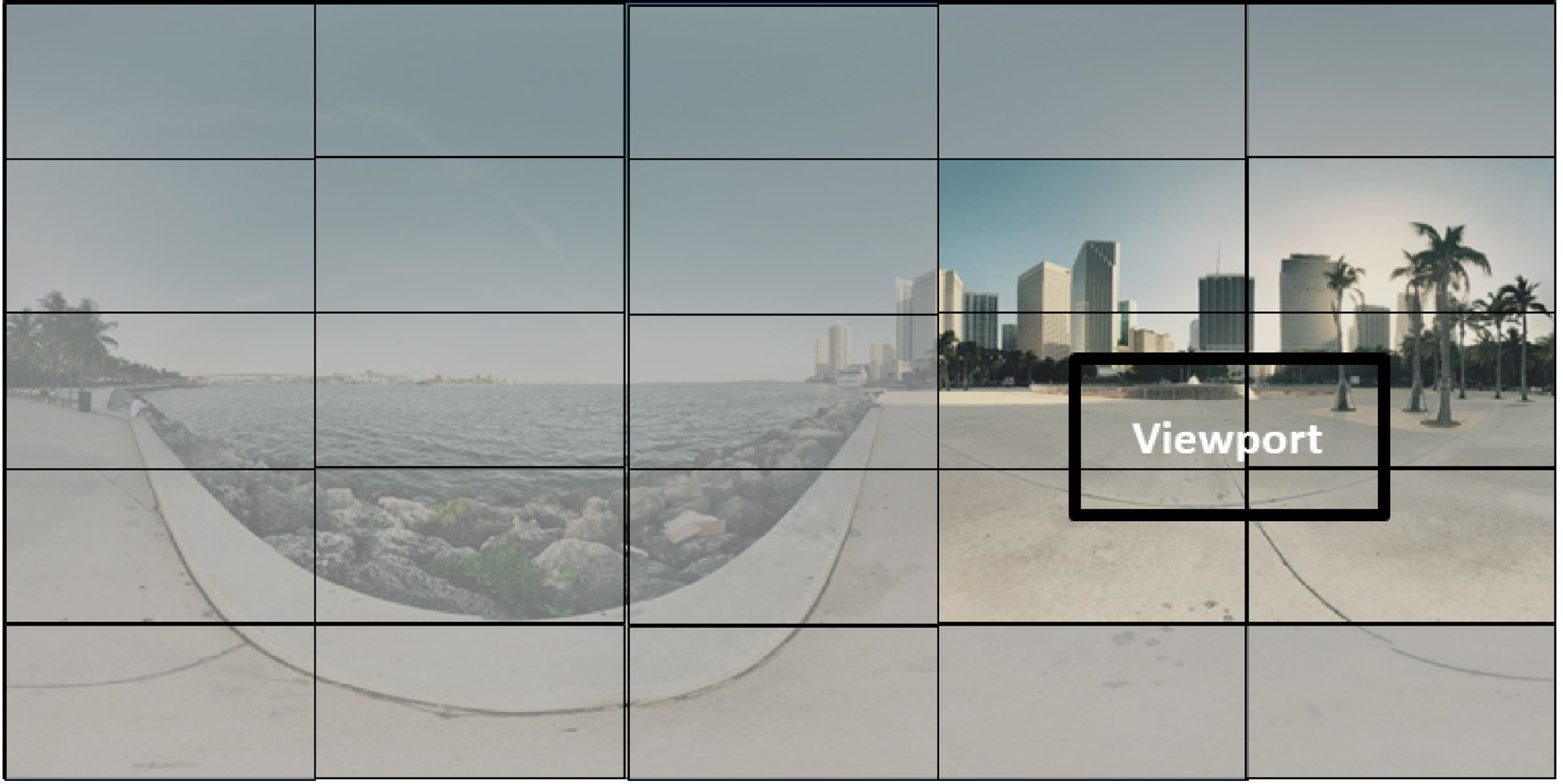}
	\caption{In tiled VR streaming, only tiles belonging to the viewport are streamed at the highest quality.}
	\label{fig:viewport-tiles}
\end{figure}

\subsubsection{Bandwidth-oriented Approaches}
\sloppy{The video content which is virtually projected to a 360° sphere can be segmented in video tiles.}
Only a limited set of tiles matches the user's viewport or FoV at a given time, as depicted on an example by Figure~\ref{fig:viewport-tiles}.
Adapting the requested quality of these tiles to the available bandwidth reduces the probability of rebuffering if the bandwidth is fluctuating.
To perform this adaptation, a throughput estimation is needed.
This can be achieved passively by measuring the throughput while receiving video data and assuming it will remain unchanged for the next transmission~\cite{petrangeli2017http, corbillon2017viewport, facebook2016next, chakareski2018viewport, xie2017360probdash, nguyen2017new}.
Upon detection of a decreasing bottleneck bandwidth,  a client reduces the requested video quality  accordingly.

\subsubsection{Prediction-based Approaches}\label{relwork:prediction-based-approaches}
Head-mounted displays (HMDs) and mobile phones  contain sensors, i.e., gyroscopes that measure the device orientation.
This data is used to determine the viewers head movement, and thus the viewport when watching 360° content.
The location and, in some cases, the movement velocity can be used to predict the future head position and the corresponding viewport.
The predicted viewports are requested in advance and buffered.
Fan et al.~\cite{fan2017fixation} and Bao et al.~\cite{bao2016shooting} use a neural network for the prediction task while
Nasrabadi et al.~\cite{nasrabadi2017adaptive}, Petrangeli et al.~\cite{petrangeli2017http}, and Xie et al.~\cite{xie2017360probdash} propose velocity-based approaches using linear regression.

\subsubsection{Popularity-oriented Approaches}
To further improve the accuracy of viewport predictions, several streaming approaches acquire additional information related to the video content.
Fan et al.~\cite{fan2017fixation} proposed identifying dynamic and visually salient regions that are likely to draw the attention of viewers.
However, a more common approach is to produce heatmaps that encode the view traces of viewers~\cite{corbillon2017360}.
Therefore, the viewing behavior of multiple users is recorded and combined.
Heated regions characterize popular areas within a video, whereas less heated regions imply less relevancy to the viewers.
This can be used to assist viewport prediction or to save these regions in a separate location or quality.


\subsection{Tiled or Monolithic}
Video segments can either be encoded in tiles, i.e., spatial segments of the video or without spatial segmentation, i.e., monolithically~\cite{corbillon2017viewport}.
In tiled approaches, each tile is encoded with a set of different bitrates, resulting in distinct quality levels.
A client can request the tiles that match its current viewport and request them in a higher quality than the tiles which are outside of the viewport, i.e., not visible to the viewer.
Monolithic approaches do not split video segments into tiles but encode them with different quality-emphasized regions (QER) as proposed by Corbillon et al.~\cite{corbillon2017viewport}.
The QER which fits the client's viewport best is requested.
The key advantage of monolithic 360° video streaming is that clients do not have to stitch the tiles back together.
However, due to the multiple representations of each video segment, redundant data tends to heavily occupy the server-side storage.

\subsection{Caching}
Caching takes advantage of storing frequently requested content at an intermediate system.
Usually caches are deployed by Content Delivery Networks (CDNs) like Akamai, Facebook CDN, or Google Global Cache.
When requesting a cached file, it is directly served from the cache, without having to fetch it from the origin, i.e., the streaming server.
It is likely that the cache-to-client connection quality is always equivalent or superior to the origin-to-client connection quality, since caches are typically placed at the network edge, i.e., close to clients and, hence, they are less prone to network congestions on the origin-client delivery path.
Obtaining data from a cache usually results in shorter RTTs, higher throughput, and thus in a faster arrival of video data~\cite{nasrabadi2017adaptive, Chakareski17}.

\subsection{Offline/Online Encoding}
Encoding video segments in an online manner, i.e., extracting the desired viewport and encoding it in desired quality upon request, is  storage-efficient.
Only the original source video needs to be stored on the server, neither tiled nor in varying quality levels.
However, this task is computationally expensive which leads to a high workload on the server's resources.
Therefore, the majority of related approaches discussed here, with the exception of \cite{nguyen2017new, bao2016viewing}, suggest to store multiple tiled representations of a 360° video, which will not be further processed by the server.

\footnotetext{Bandwidth adaptation, as in adjusting quality demands based on available bandwidth, is only used when transitions are disabled.}

\section{360° HMD Video Player}\label{existing-software}
\begin{figure}[h]
	\centering
	\includegraphics[width=1\linewidth]{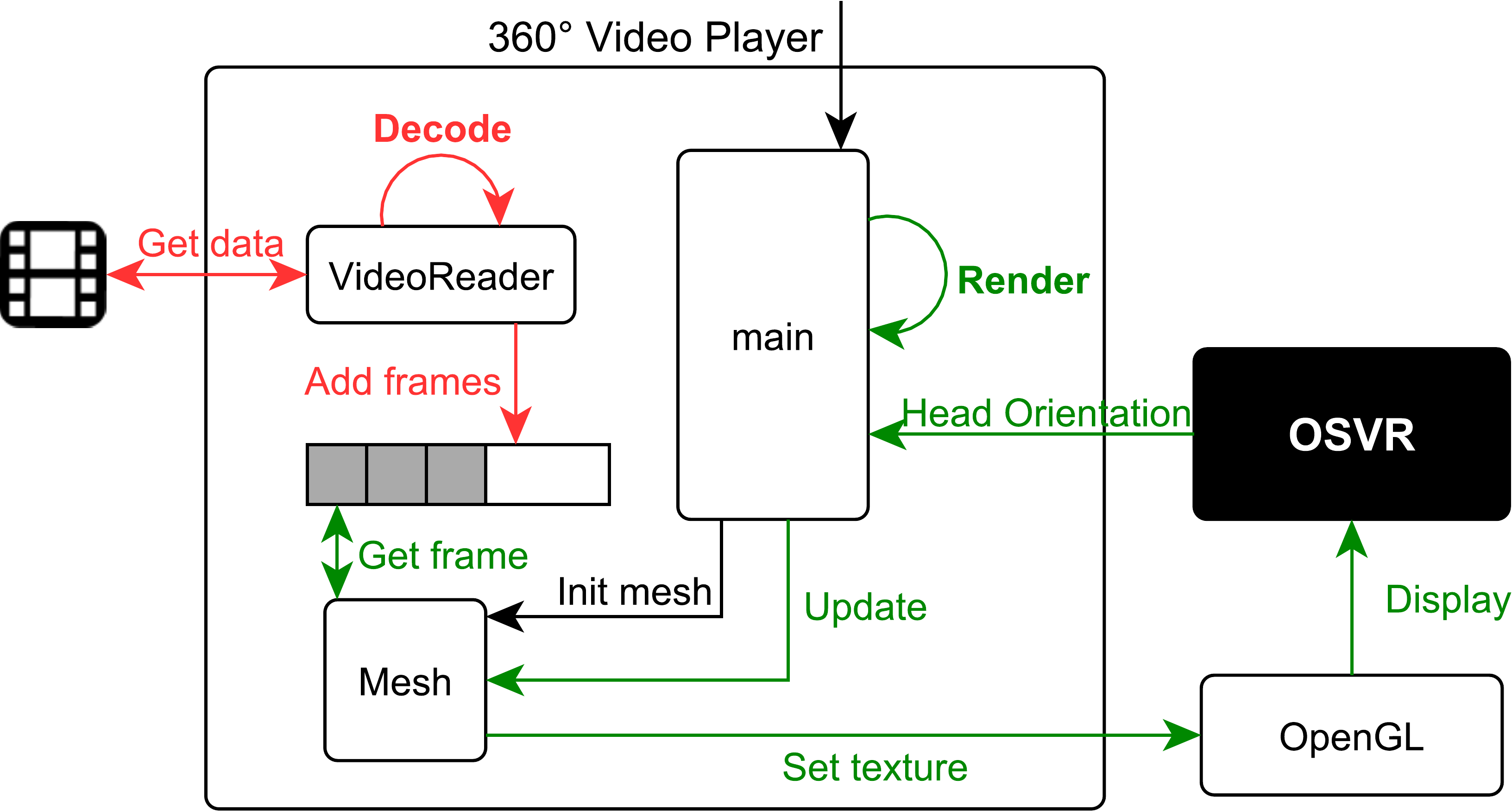}
	\caption{Architectural overview of the used system components proposed by Corbillon et al.~\cite{corbillon2017360}}
	\label{fig:basis}
\end{figure}

Our system is based on the software from Corbillon et al.~\cite{corbillon2017360}.
It is publicly available on GitHub~\cite{36player}. 
The software was built for tracking the head movement of several participants while watching various 360° videos.
For this purpose, it is capable of rendering locally saved 360° videos to a HMD.
The authors use the OSVR HDK2\footnote{\url{http://www.osvr.org/hdk2.html} [Accessed: \today]} HMD for their project \cite{corbillon2017360}.
Figure~\ref{fig:basis} provides a high-level overview of the software components.
The video player consists of two fundamental processes.
Their purposes are to decode a local video file and render it to the HDK2, respectively.
Figure~\ref{fig:basis} colors their individual steps in red (decoding) and green (rendering).

\noindent In the following, we introduce the key components of the architectural overview presented in Figure~\ref{fig:basis}.

The \textbf{main} component is the entry point of the application. When launched, it is provided with a configuration file, containing the path of the locally saved video file.
Next, \emph{Main} initializes other components and starts the rendering process.

\textbf{Mesh} maps equirectangular frames (i.e., each frame has the same dimensions) on the inner surface of a three-dimensional body.
The camera is placed in the center of that body, i.e., a cube in this case.
A more intuitive choice would be a sphere, as this is the geometric body that the captured 360° video is projected on in the first place.
However, in this scenario, the resulting visual effect using a cube or a sphere are indistinguishable to the viewer~\cite{harrison2003basic}.

The \textbf{VideoReader} continuously reads encoded video data from a local file.
This data is decoded, resulting in raw frames which are added to an internal frame buffer.

The \textbf{OSVR} black box is an abstraction of all OSVR components that are included in the system.
For this paper, only the OSVR HDK2 HMD is important.
It provides the video player with orientation information and displays the video.

\textbf{OpenGL} is the graphics API used. It interacts with the GPU to render 2D and 3D scenes to the screen.

The decoding and rendering processes are isolated from each other for two reasons:
(i) Running them in parallel on multiple CPU cores is faster than sequentially decoding and rendering on a single core.
(ii) The refresh rate of the HMD's display is usually not equal to the video's frame rate.
The HDK2's display has a refresh rate of 90 Hz.
It is required to update the displayed image accordingly to avoid perceivable stuttering, which is done in the rendering process. However, most videos do not provide this frame rate, so when a render update occurs, a new video frame cannot always be supplied.
Fortunately, the current head orientation can still be applied to the last displayed video frame, resulting in an appropriate rendered image.
Thereby, the updates of the current video frame and current rendered image have to happen desynchronized.
For this reason, \emph{VideoReader} stores decoded frames in a buffer, so that the \emph{Mesh} component can pick the appropriate frame according to the current playback timestamp or keep the last displayed frame if needed.

The render update instruction is issued from the \emph{main} component.
For each update a head orientation is provided by \emph{OSVR}.
\emph{Mesh} then potentially gets a new frame from the internal frame buffer and performs the viewport extraction using the orientation information conveyed by \emph{main}. \emph{OpenGL} receives this as a texture and prompts the screen rendering.

\section{System Design}\label{sec:system-design}
This section discusses the design of our proposed system.
We propose transitioning between an approach of \emph{viewport prediction-based quality adaptation} in case of sufficient bandwidth and a \emph{tile-popularity-based approach} in case the bandwidth is insufficient to deliver the tiles requested by the first adaptation approach.
Thereby, we actively overcome bandwidth fluctuations and shortages, since we deliver popular content from a cache close to the user which does not suffer from the bandwidth impairments between client and content origin.
Specifically, we assume a cache on the delivery path that is beneficial in case of bandwidth impairments between the cache and the content origin.
Examples for such impairments are a high load in the broadband access ISP's core network, network congestion between the ISP of the user and the ISP of the content origin, or non-optimally configured traffic shaping.
To achieve the aforementioned transition-capabilities, we extend the design presented in Section~\ref{existing-software} as follows (ref. Figure~\ref{fig:overview}):
\begin{figure}[t]
	\centering
	\includegraphics[width=1\linewidth]{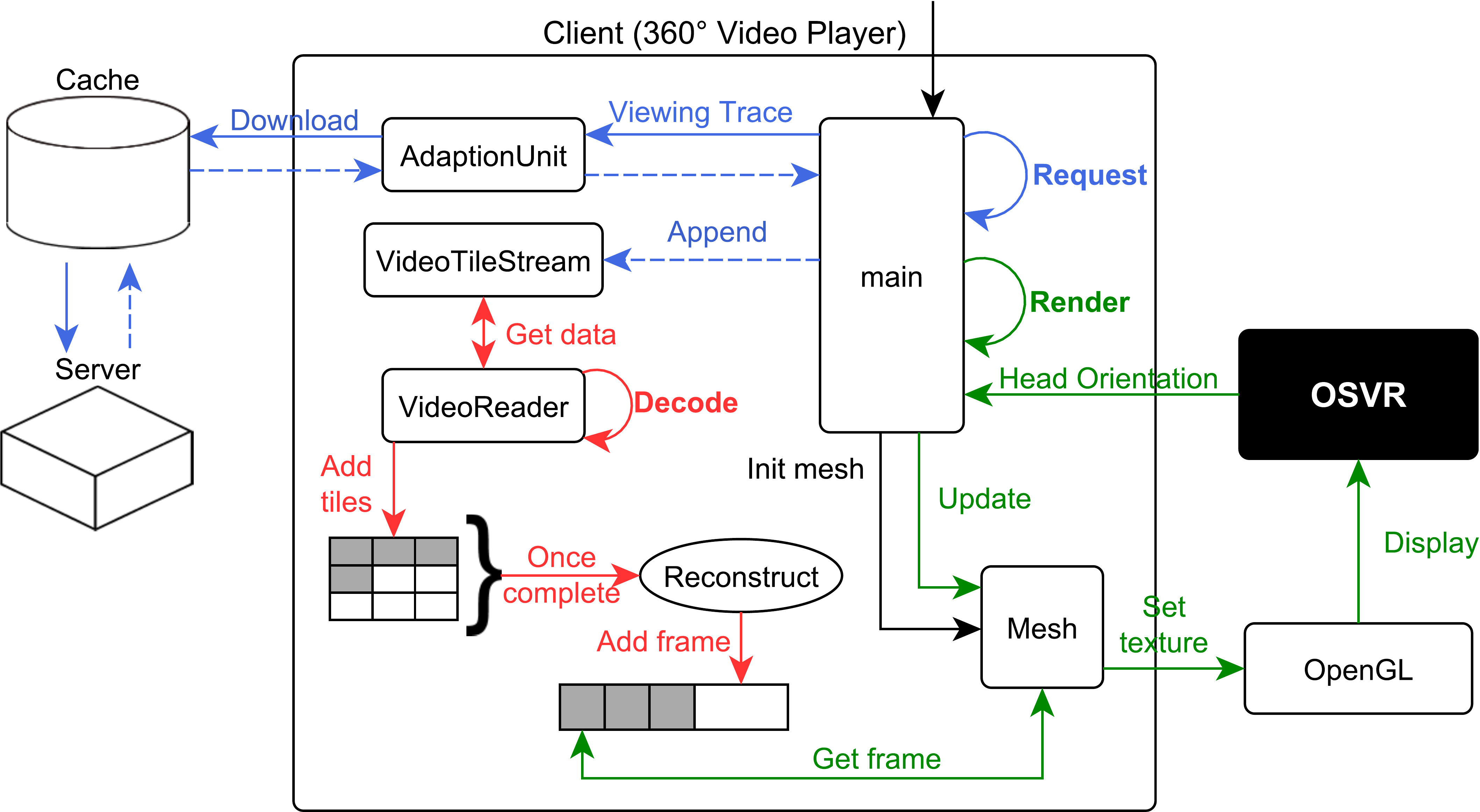}
	\caption{System architecture, each color refers to a process running in parallel.}
	\label{fig:overview}
\end{figure}

\begin{itemize}
	
	\item \textbf{Network streaming:} The video tiles are regularly requested from a remote server that is reachable by the client.
	In addition, an intermediate cache is placed on the delivery path, between the content server and the client.
	
	\item \textbf{Spatially tiled videos:} The \emph{VideoReader} is extended to receive and stitch tiled video segments based on the  video manifest, i.e., the Media Presentation Description (MPD) file which it requests from the server during the initialization phase.
	The MPD contains information about the requested video's segments, e.g., their duration, location, and the bandwidth required to download them sufficiently fast.
	Additionally, the MPEG-SRD extension is used to describe the spatial relationship between individual tiles~\cite{niamut2016mpeg}.
	The \emph{AdaptationUnit} determines the video quality for all tiles and downloads them.
	Therefore, it keeps track of the most recent user head orientations.
	The \emph{VideoReader} receives its video data from a new component, i.e., the \emph{VideoTileStream} since we no longer assume a monolithic video.
	The tile-based decoding and stitching is performed by the \emph{VideoReader} which creates entire coherent video frames.
	
	\item \textbf{Quality adaptation:} A component is introduced that performs the quality adaptations based on sensory or content-based predictions and performs transitions between these quality adaptation mechanisms based on the measured available network bandwidth.
	
\end{itemize}

When the client application is started, a connection to the cache system is established.
Client requests are potentially forwarded to the server, depending on whether the requested content is cached.

In a next step, the video manifest file, i.e., the MPD is downloaded.
The \emph{AdaptionUnit} component uses this information to download individual tiles.
To decide in which quality each tile should be requested, the \emph{AdaptionUnit} performs quality adaptations based on the viewing trace, i.e., the latest user head orientations, available bandwidth, and tile popularity.
Information about the tile popularity is also contained in the MPD file (cf. Section~\ref{sec:sys-popularity}).
While new encoded video segments are arriving, the decoding process simultaneously takes care of decoding the video data.
The decoded tiles are reconstructed and added to the same internal frame buffer that is used in the original software (cf. Section~\ref{existing-software}).
Our software is publicly available for others to reproduce our results and for further research projects~\footnote{\url{https://github.com/arizk/360transitions} [Accessed: \today]}.

\subsection{Transitions}
\begin{figure}[h]
	\centering
	\includegraphics[width=1\linewidth]{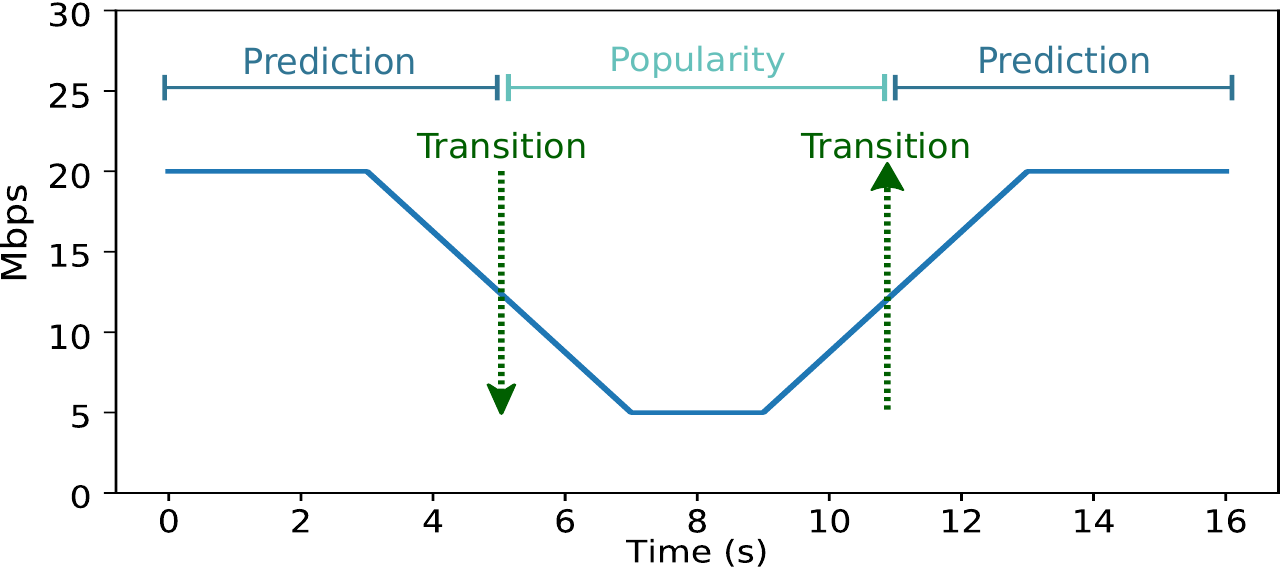}
	\caption[]{The continuous line represents a network bandwidth trace. When it drops below the transition threshold $A$, a transition to popularity based adaptation occurs. If the bandwidth recovers, we transition back to prediction based adaptation.}
	\label{fig:transition}
\end{figure}

Both content-based and sensory prediction mechanisms have their benefits and drawbacks, depending on, e.g., the available bandwidth.
Thus, combining them is likely to result in an improved QoE.
Sensory prediction provides the viewers with high-quality video areas depending on their head movement.
However, this approach is prone to prediction errors and may result in either stalling events or low video quality when the server-to-client connection's bottleneck bandwidth declines.
Content-based prediction might still provide high-quality video, even in the case of throughput decline.
In this context, an intermediate cache will store popular video areas, as they are frequently requested.
Although allowing for persistent video quality, it is unlikely that a viewer will only look at popular video regions for the entire duration of the video.
Therefore, solely using content-based prediction is not an ideal solution.
Related approaches have combined both adaptation methods by complementing viewport prediction with different kinds of content-based mechanisms, e.g., Fang et al.~\cite{fan2017fixation} and Qian et al.~\cite{qian2016optimizing}.
While this does improve the video quality in an ideal network environment, viewers will still experience a low video quality or even stalling as soon as bandwidth demands cannot be met by the network.
To cope with this issue, we introduce transitions between these mechanisms.
Hence, either sensory or content-based prediction is used at a given time.
Transitions do not occur randomly, but get triggered by a measurable metric within in the streaming system.
For instance, if the computational power of the video playback device is not high enough to perform viewport predictions  sufficiently fast, only a content-based prediction mechanism should be used.
This might be useful for complex viewport prediction methods that involve, e.g., the training of a Neural Network, or in case of high CPU workload caused by other processes.
In our approach, we make use of the estimated bandwidth metric to determine a transition.
We consider the estimated bandwidth as the most important contextual information.
If the bandwidth estimate drops below the transition threshold $A$, a transition from sensory to content-based prediction occurs.
Note that we estimate only the bandwidth between client and content origin, i.e., of non-cached video content.
In case $A$ is underrun, popular tiles are requested in high quality.
Thereby, mainly popular content from an intermediary cache is retrieved, avoiding the slow connection to the content origin.
Eventually, when the connection quality recovers, a transition back to sensory prediction occurs.
Figure~\ref{fig:transition} illustrates this concept.
Since we do not abort started downloads of video tiles, a transition can only occur between two video segments, i.e., at most once per second given a video segment length of 1\,s.
Note that a hysteresis technique can be employed to limit the transition frequency.

\subsection{Viewport Prediction}
The first step to predict a future viewport is estimating the future head orientation.
The \emph{main} component provides the \emph{AdaptionUnit} with a viewing trace every time a viewport prediction is performed.
This viewing trace contains the latest head orientations of the viewer.
Each entry consists of a timestamp and its corresponding head orientation.
Using the latest entries, a linear regression model is fitted.
The number of time points contained in the trace  is a configurable parameter.
Alternatively, more complex regression methods could lead to more accurate prediction results~\cite{bao2016shooting}.
Note however that the accuracy penalty using linear regression is negligible here~\cite{bao2016shooting}.
The regression model takes a timestamp $t$ as an input and outputs an estimate for the viewer's head orientation at time $t+x$.
We compute the visibility of each tile inside the viewport corresponding to the predicted head orientation at $t+x$.
Finally, tiles are sorted by their respective visibility scores and are assigned matching quality levels.
While assigning quality levels, the increasing bandwidth requirement is continuously compared to the bandwidth estimate.
As soon as the estimate is surpassed, adaptation is stopped and the last modified tile quality level is reset.
This way, we prevent requesting too much data beforehand, as this may lead to video playback stalling.

\subsection{Popularity}\label{sec:sys-popularity}
\begin{figure}
	\includegraphics[width=0.75\linewidth]{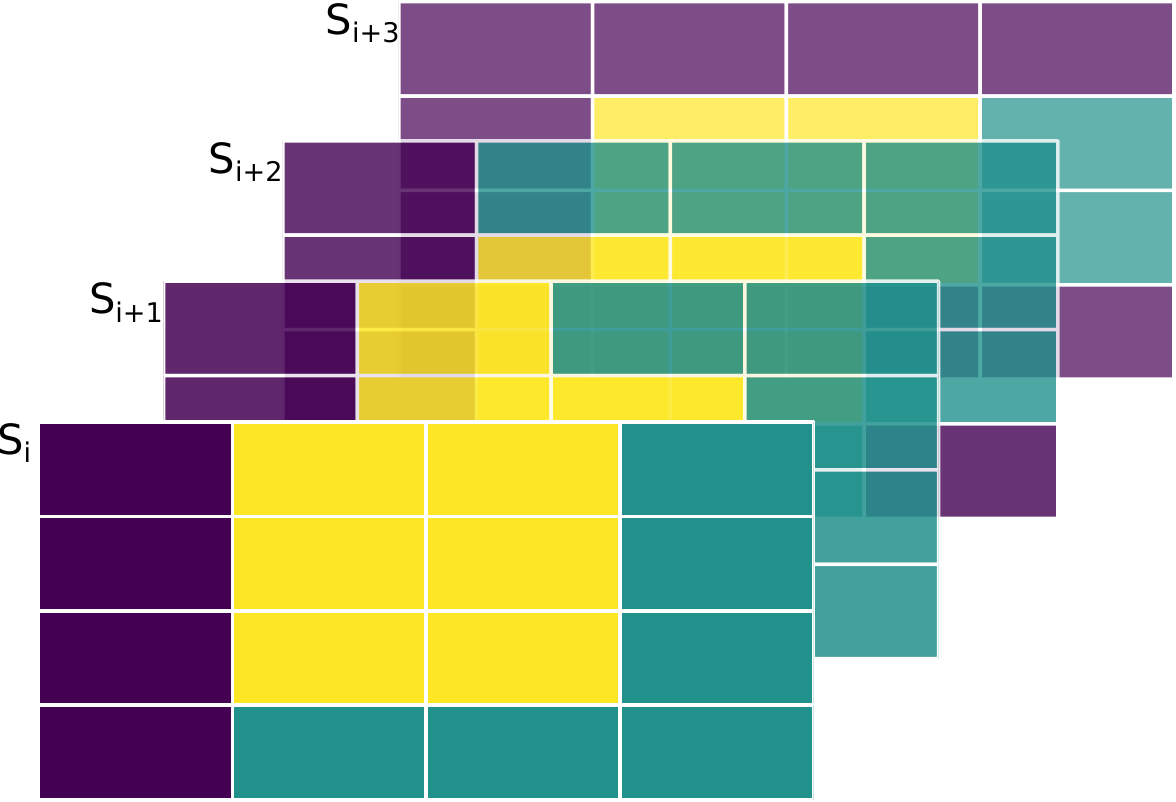}
	\caption{Tile popularity distribution in consecutive video segments $S$ with time index $i$. A brighter color refers to a higher popularity (yellow, green, purple). } \label{fig:poptrace}
\end{figure}
The tile popularity is only involved in determining the tile quality levels when a transition due to low bandwidth has occurred.
Note that the \emph{AdaptionUnit} is not responsible for identifying popular video areas.
It rather utilizes popularity information included in the manifest file when needed.
The manifest file contains data about tile popularity for each video segment.
Figure~\ref{fig:poptrace} depicts the tile quality distribution for four video segments ($S_i, ..., S_{i+3}$) with a regular $4\times 4$ tiling grid.
We can observe that, especially in segment $S_{i}$, the center and upper area appears to be the most popular video region.
The quality levels displayed in this figure are directly taken from the popularity information in the corresponding manifest file.
Thus, internally, tile popularity is represented by a distinct quality level per tile and segment.
In the following, we refer to this as the \emph{popularity trace}.
This makes it simple for the \emph{AdaptionUnit} to consider the manifest's popularity information.
Tile quality levels for a segment are fetched from the popularity trace and downloaded accordingly.
If the popularity quality adaptation is active, the viewer's movements will not have an effect on the downloaded tile qualities.
Thus, in order to see high-quality video, a viewer is \emph{guided} to the popularity trace.
This approach can be seen a natural guidance to get the users to watch the most popular parts of the video.
An alternative approach is to use bitrate adaptation, which reduces the overall quality of an entire video segment and hence does not guarantee that the popular parts of the video are displayed in high quality.


In our evaluation, popularity information is obtained from~\cite{corbillon2017360}, containing head movement traces of 59 participants recorded while they were watching various 360° videos.
Therefore, we can only compute popularity information about the videos that were shown in this experiment.
We imagine, however, that user feedback or automated saliency calculation methods can be used to automatically generate such heatmaps.
We also note that the corresponding segments are likely to be at the edge caches due to their high popularity.

\section{Evaluation}\label{sec:eval}
This chapter presents the evaluation of our 360° video streaming system.
The goal of the evaluation is to show the benefits and drawbacks of our system compared to alternative approaches.
Therefore, we show the effects of transitions between quality adaptation mechanisms on bandwidth usage, video quality, and the resulting QoE, i.e., by measuring playback stalling events besides the visible video quality.
As a benchmark, the naive streaming approach of not incorporating viewer head movement into quality selection is chosen.
A comparison of our approach with naive streaming is important, as this is how major VoD platforms implement streaming of omnidirectional content~\cite{graf2017towards}.
Thereby, it supports comparing our approach with the state-of-the-art~\cite{graf2017towards,zare2016hevc,bao2016shooting}.

We use a trace-driven simulation for our experiments.
Therefore, the 360° video player (cf. Section~\ref{existing-software}) was narrowed down to only execute the steps necessary for collecting the required data.
Hence, decoding and rendering functionalities were skipped.
Our simulator uses the viewing traces from~\cite{corbillon2017360} to simulate viewer head movements.
Additionally, network traces collected by Winstein et al.~\cite{winstein2013stochastic} are used to throttle server throughput. They contain timestamps representing the arrival time of 1500-byte packets in various cellular networks over several minutes.

\subsection{360° Video Dataset}\label{subsec:360-dataset}
\begin{figure}[t]
	\includegraphics[width=1\linewidth]{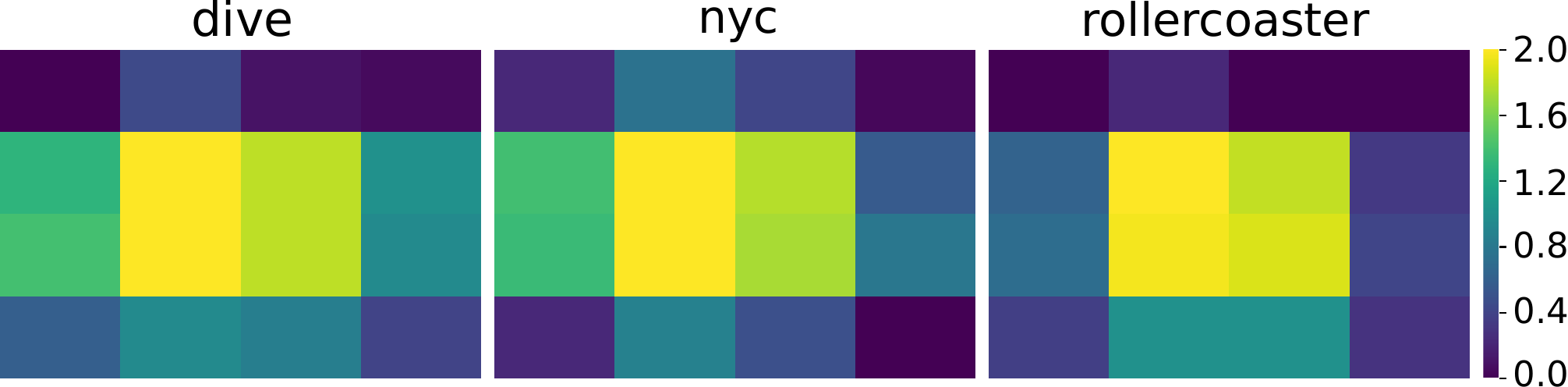}
	\caption[Average tile quality of video popularity traces.]{Average tile quality of the used video popularity traces for three different qualities. Brighter $\equiv$ more popular.
		The videos are encoded in three quality levels. Here, 0 represents lowest quality. In all videos, the center area (front) appears to be the most popular area. This is mainly because when turning from one side to the other, viewers will most likely not perform a full turn, but rather first turn their head to the front and then to the other side, as this mimics their natural head movement. However, with \emph{rollercoaster} a more distinct quality distribution is observable, with the main focus point being the center area. This is expected, since the roller coaster's rail is visible in the front, which is what most viewers are interested in viewing.
	}\label{fig:poptraceavg}
\end{figure}

We choose three videos from the dataset provided by~\cite{corbillon2017360} with content that motivates different viewer movement: \emph{dive}, \emph{nyc}, and \emph{rollercoaster}.
In the \emph{rollercoaster} video, viewers mainly look at the roller coaster's track, while the other videos contain no particular focus point, and thus, no specific movement can be expected.
This characteristic is reflected in Figure~\ref{fig:poptraceavg} by a higher popularity of the leftmost and rightmost tiles for \emph{dive} and \emph{nyc} compared with the \emph{rollercoaster} video.
Table~\ref{tbl:videos} provides an overview of the videos' key characteristics.

\begin{table*}[t]
	\small
	\caption[Overview of the videos used for evaluation.]{\textbf{Overview of the videos used for evaluation.} The ``Original Size'' column displays the videos' sizes before they were tiled and segmented. The ``Tiled Size'' columns contain the accumulated sizes of all tiles belonging to a quality level, per video. The \emph{high} quality level matches the original video's bitrate. Medium and low quality versions were encoded with 0.25 and 0.0625 times the original bitrate, respectively. }
	\begin{tabular}{l X c c c c c c c}
		\toprule
		Video & Content & \rotatebox{25}{Duration}   & \rotatebox{25}{Original Size}   & \rotatebox{25}{Tiled Size High }  & \rotatebox{25}{Tiled Size Medium}    & \rotatebox{25}{Tiled Size Low }    \\
		\midrule
		dive & \makecell{Diving scene with a slowly moving\\ camera and no main focus point.} & 30 & 73.69 & 80.24 & 17.87 & 4.52  \\
		
		nyc & \makecell{Timelapse with static camera and fast\\ moving objects. No main focus point.}  &40 & 67.57 & 109.82 & 28.21 & 6.64  \\
		
		\makecell{roller-\\coaster} & \makecell{Fast moving camera attached to\\ roller coaster. Main focus on rail.} &40 & 104.61 & 116.21 & 34.44 & 9.25  \\
		\bottomrule
	\end{tabular}
	\label{tbl:videos}
\end{table*}

For preprocessing the videos, we transcode them with factors of 1, 0.25, and 0.0625 times the original bitrate, so that three quality levels exist per video.
The tiled videos are temporally segmented into 1.5\,s parts since typical segment lengths range between 1\,s and 2\,s~\cite{xu2018probabilistic,qian2016optimizing}.
Decreasing the segment length will lead to faster quality adaptations to changing head orientations, as the tiles are updated more frequently.
However, shorter segment lengths may be more prone to sudden throughput changes, as they have to be downloaded in a very short time.
With much longer segment lengths, viewport prediction becomes very inaccurate, up to a point to where future head orientations are practically unpredictable.
After preprocessing the videos, popularity information is generated, using the viewing traces available for each video.
This information is added to the videos' respective MPD manifest files.
Finally, the manifest and video files are stored on a HTTP server's file directory.

\subsection{Preprocessing}
To prepare a 360° video for our streaming system, it has to be spatially tiled and temporally segmented.
Finally, a  {DASH} manifest file has to be generated.
We use \emph{ffmpeg}\footnote{\url{https://www.ffmpeg.org/} [Accessed: \today]} and \emph{MP4Box}\footnote{\url{https://gpac.wp.imt.fr/mp4box/} [Accessed: \today]} to accomplish that.
We first crop a video in a fixed but arbitrary $n\times m$ grid pattern to generate the tiles.
The cropped videos are saved in a fixed but arbitrary number of versions $q$, all of them differing in bitrate.
Each tiled video of length $d$ is then temporally segmented, with an a segment length $s$.
The {DASH} standard introduces an initialization segment per video stream.
In total, including the manifest file, the preprocessing application generates $n \cdot m \cdot q \cdot \left(\left\lceil \frac{d}{s} \right\rceil+1\right)+1$ files.
For a 40\,s video and a $4\times 4$ tiling pattern, three quality levels (cf. Section~\ref{subsec:360-dataset}), and 1.5\,s segment length, this amounts to 1345 files.
Having to store this many files for one single video can be avoided by using byte ranges.
Rather than saving temporal segments to individual files, using {DASH} byte ranges, they are grouped in one coherent file.
In this case, the manifest file contains byte ranges for each segment~\cite{stockhammer2011dynamic}.

\subsection{Caching}

\subsubsection*{Cache Initialization}
For most experiments, we use an intermediary cache between the server and the client.
Specifically, we assume a telco-owned rentable cache from which we use a dedicated cache storage area per video.
To construct a realistic steady-state environment, we initialize the cache by first generating a random permutation of a set of 30 randomly chosen viewing traces from \cite{corbillon2017360} belonging to the video that is currently used for evaluation.
Note, this is the initial state at the simulation start but the cached content can change during the simulation, e.g., by a changing tile popularity.
In a productive environment, the cached content caches dynamically depending on the requested video tiles.
However, we assume the content to converge to a steady state soon after the first users have watched the video.
Each viewing trace is then sequentially used in a viewing simulation, which downloads the tiles in qualities  corresponding to their visibility depending on the head orientations contained in a trace.
The tiles in the center of the FoV are streamed in the highest quality, while the non-visible tiles are streamed in the lowest available quality (cf. Section~\ref{subsec:360-dataset}).

\begin{figure}[b]
	\includegraphics[width=1\linewidth]{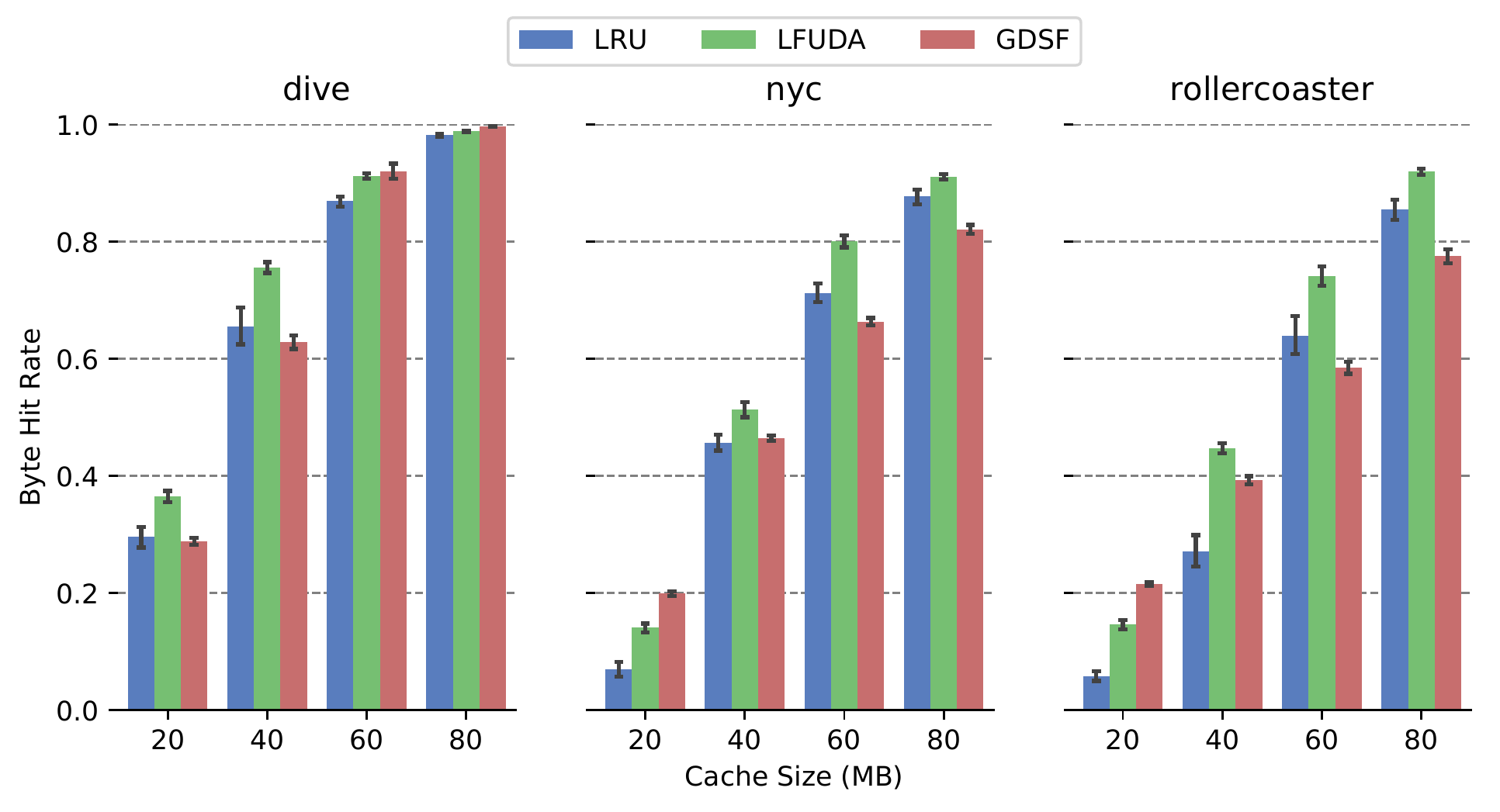}
	\caption{Byte Hit Rate (BHR) when downloading the popularity trace of different videos, averaged over 30 different experiments with 95\% confidence intervals.}
	\label{fig:eval:replacement_chr_bhr}
\end{figure}

\subsubsection*{Cache Replacement Policies}
When a cache reaches its capacity limit, a policy is needed that decides which files are evicted to create space for new requested content to be cached.
To this end, we evaluate each policy by measuring Cache Hit Rate (CHR) and Byte Hit Rate (BHR)  when downloading a video's popularity trace.
This represents how the cache contents match our popularity trace.
We consider three replacement policies\footnote{\url{http://www.squid-cache.org/Doc/config/cache_replacement_policy/} [Accessed: \today]}:
\begin{itemize}
	\item {Least Recently Used (LRU)}: Keeps the recently requested content in the cache.
	\item {Least Frequently Used with Dynamic Aging (LFUDA)}: Keeps popular contents in the cache, regardless of its size, and thus optimizes BHR at the expense of CHR since one large, popular object will prevent many smaller, slightly less popular objects from being cached.
	\item {Greedy-dual Size Frequency (GDSF)}: Optimizes the CHR by keeping smaller popular contents in the cache, so it has a higher CHR probability. It achieves a lower BHR than LFUDA though, since it evicts larger (possibly popular) contents.
\end{itemize}
First, we evaluate which cache eviction policy works best in conjunction with our system using tile popularity-based quality adaptation.
Furthermore, as cache capacity has  an impact on how each policy performs, we repeat this process for a capacity range $C \in \{20, 40, 60, 80\}$\,MB.
These cache sizes were chosen with regard to the sizes of our preprocessed videos.
Hence, we mimic real-world caches which store on average a particular share of the video content.
For example, the cache can store from $12.5\%$ up to $50\%$ of the 160 MB \emph{rollercoaster} video.
This helps us to make assertions on how cache capacity settings should be determined based on individual video sizes.

To specify squid's replacement policy and cache capacity, the \emph{cache\_replacement\_policy} and \emph{cache\_dir} configuration directives are used inside the \emph{squid.conf} configuration file, respectively.
For each policy and cache size, \emph{squid.conf} is modified and squid is restarted to load the modified settings.
Then, the cache is initialized using the initialization process described in the previous section.
We measure CHR and BHR when downloading the popularity trace of the current video by checking the \emph{X-Cache} HTTP response header entry.
As the cache varies in content after initialization, this process is repeated 30 times.
For each of these iterations, the cache is initialized with a new viewing trace permutation. Figure~\ref{fig:eval:replacement_chr_bhr} depicts the average BHR measured over 30 iterations for varying cache sizes.
Overall, LFUDA achieves the best BHR, as the popularity of the entire history of viewing traces is considered, instead of just the recently downloaded ones, e.g., when using LRU.

For the \emph{dive} video and cache sizes 60\,MB and 80\,MB, GDSF also achieves a high BHR.
Note that GDSF tries to optimize CHR by preferring to cache smaller files.
Consequently, a high CHR can be reached at arbitrary cache sizes by emphasizing to store all low quality tiles before considering popular high-quality tiles.
However, we argue that the BHR is a more valuable metric since it directly reflects the amount of data traffic saved.
LRU does not perform well since it only adapts the cache status to recently requested files.
Based on these results, LFUDA is our choice for the following evaluations.


\subsection{Prediction Accuracy}
\begin{figure}[b]
	\includegraphics[width=1\linewidth]{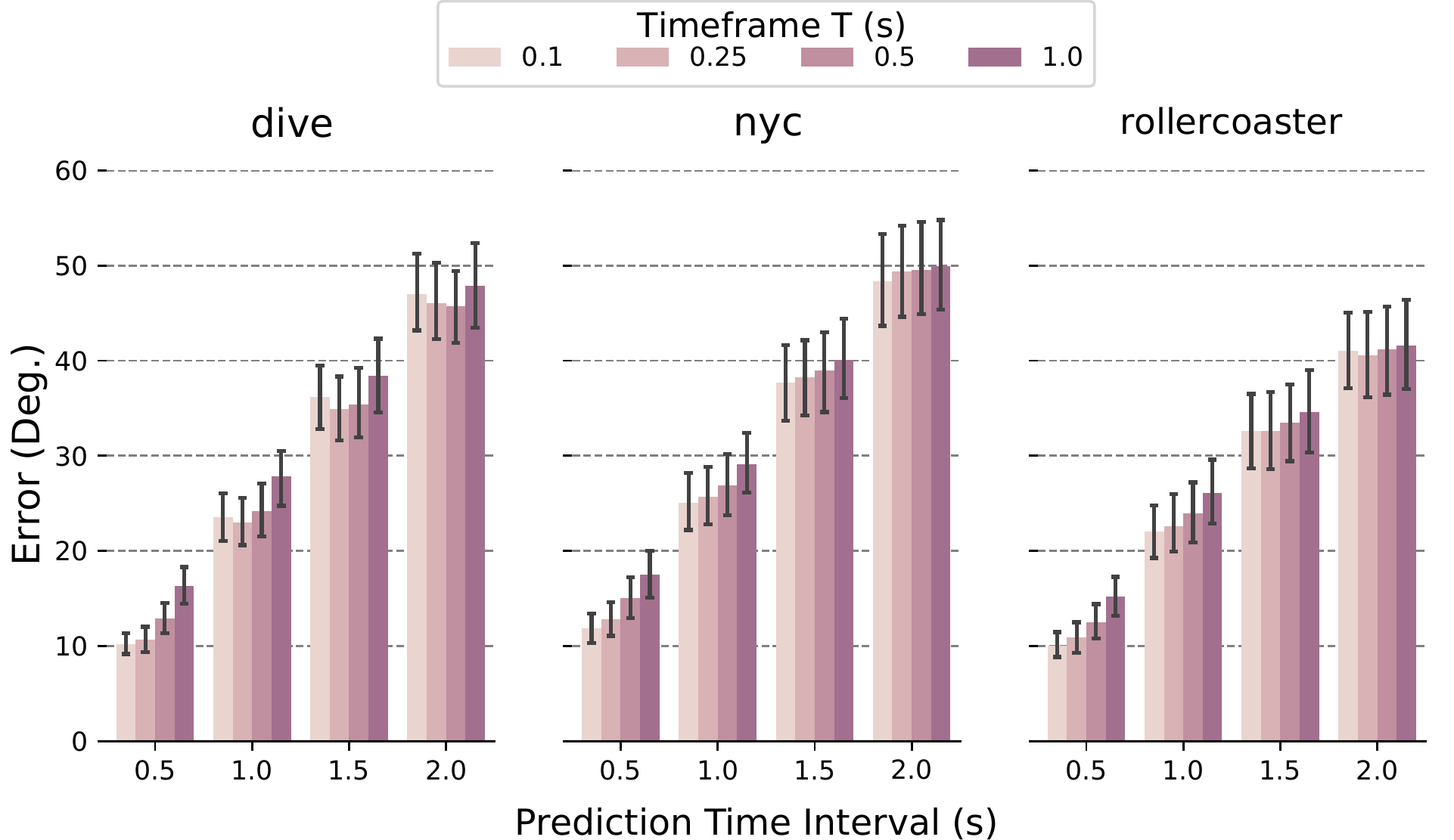}
	\caption{Prediction errors depending on prediction time $I$ and timeframe $T$ used to compute the linear regression model.}
	\label{fig:eval:prediction}
\end{figure}
An accurate viewport prediction is needed to provide the viewer with high-quality tiles within the FoV.
We evaluate the prediction accuracy by measuring the prediction error.
This is done for each prediction interval $I \in \{0.5, 1.0, 1.5, 2.0\}$\, seconds.
For the predictions, we use a linear regression model similar to Qian et al.~\cite{qian2016optimizing} and Xu et al.~\cite{xu2018probabilistic}.
Further, we choose the prediction interval to be equal to the video segment length.
Hence, the results imply how viewport prediction would perform if an equivalent segment duration is chosen.
Furthermore, we evaluate which time frame length $T$ should be chosen to generate an accurate linear regression model.
When $T=1$, every head rotation sample collected within the last second is used to compute the regression model, i.e., between 40 and 180 samples per second.
Depending on the amount of samples, i.e., the time frames chosen, the prediction accuracy might vary.
As we use pre-recorded viewing traces, the prediction error can be computed by comparing a predicted head orientation with the actual head orientation contained in the trace at the according timestamp.
We represent the \emph{difference} between two orientations using their orthodromic distance, defined as the shortest distance on the surface of a sphere.
We measure it in degrees, so the maximum distance between two points is 180°.
The data acquisition is done by iterating through a viewing trace in $s$ second steps, with $s$ being the video segment length.
In each iteration, the head orientations of the last $T$ seconds are collected and then passed to the prediction model, which computes a head orientation for a future timestamp after $I$ seconds.
Next, the orthodromic distance between the predicted and actual orientation is computed.
To get an average prediction error per (I, T, video) tuple, all viewing traces available for a video are combined.


Figure~\ref{fig:eval:prediction} presents our results.
We observe that the prediction error correlates with the prediction interval.
The error increases the further into the future a prediction is performed.
This is expected, as viewers are likely to change their head movement direction, and thus, deviating from the linear regression model's prediction.
Further, the average errors computed on the \emph{rollercoaster} video are noticeably lower than on the other two videos since it has a distinct point of interest.
Thus, it is likely that viewers will not turn their heads as much as in other videos, making it easier to predict a future orientation.
Finally, the average errors created by varying timeframes most often do not significantly differ.
Only with a prediction interval of 0.5\,s, the 0.1\,s timeframe consistently yields a significantly superior result to the 1\,s timeframe.
As a timeframe of 0.1\,s always leads to an equal or lower prediction error compared with larger prediction time invervals, it should be used for prediction.
Additionally, the computational effort involved in fitting the regression model with fewer samples is lower.
Note that a prediction error as low as 10°-20° can be easily mitigated by requesting an area that is slightly larger than the viewport itself~\cite{qian2016optimizing}.
However, as we can see from the evaluation results, prediction intervals of 1.5\,s and longer lead to much higher prediction errors.
Thus, ideally, segment lengths of 1\,s and lower should be chosen to guarantee high prediction accuracy.

\begin{figure}[b]
	\centering
	\includegraphics[width=0.9\linewidth]{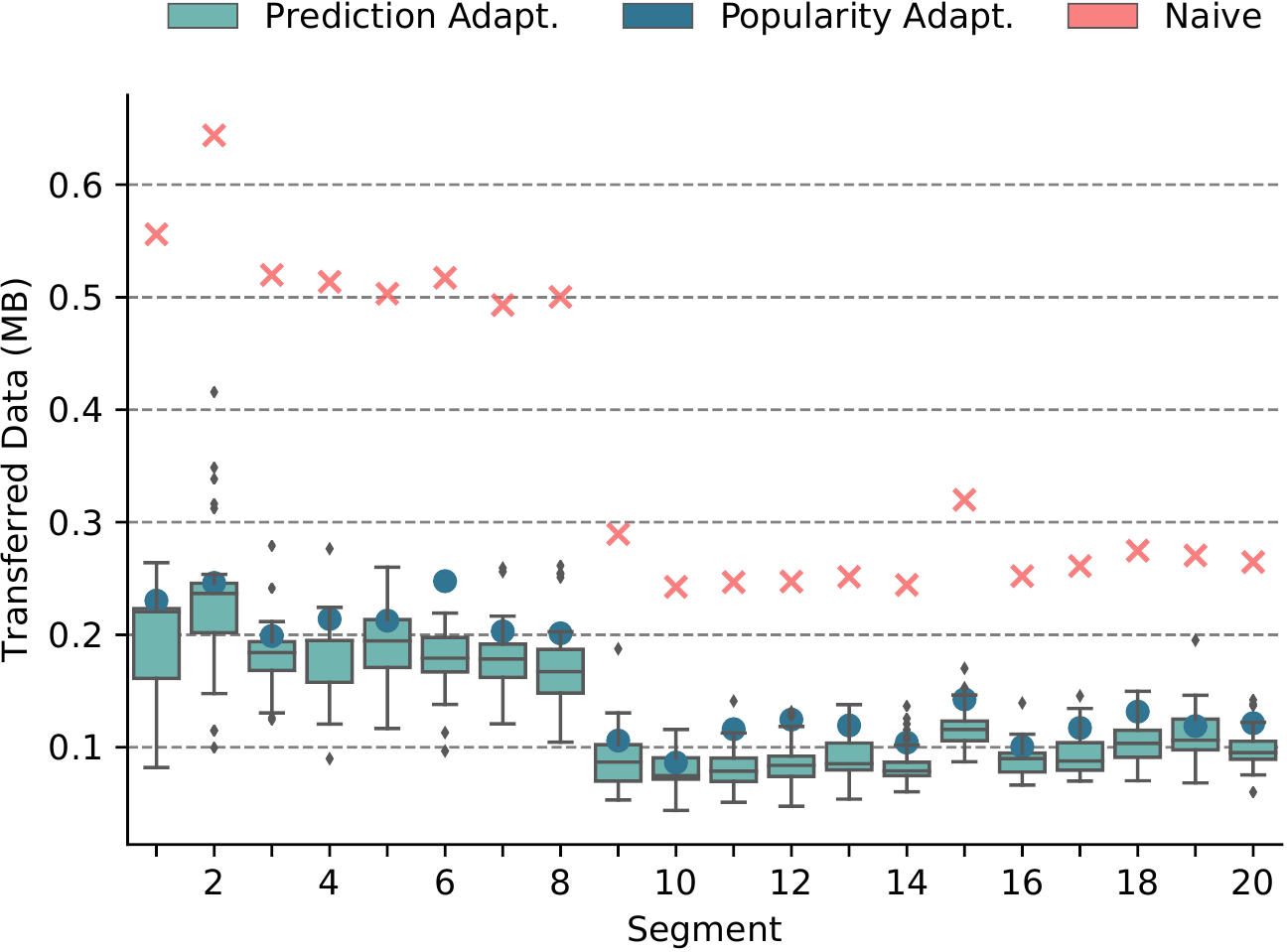}
	\caption{Bandwidth usage of multiple adaptation strategies while downloading the \emph{dive} video.}
	\label{fig:eval:bandwidth_diving}
\end{figure}

\subsection{Bandwidth Usage}\label{eval:bandwidth}
Transferring only a set of video tiles in high quality lowers the bandwidth required for video streaming compared to streaming the entire video in high quality.
In this section, we evaluate the effects on bandwidth usage when only transferring sub-parts of 360° videos in the highest quality.
To measure the bandwidth used, the file size of every downloaded tile stream segment is considered.
Thus, the results depend on the quality levels in which the videos are encoded.
For data acquisition, we streamed each video multiple times and measured the amounts of received bytes as explained in detail in the following.
First, all segments are downloaded in highest quality to represent the naive approach.
For the naive approach, we also benefit from cached content, however, on a small hit rate compared to other video quality selection approaches.
Next, only the video tiles captured by the tile popularity trace belonging to the current video are downloaded. 
Finally, 30 viewing traces are used to stream the video 30 times using viewport prediction.
As our transitioning scheme works by using either popularity or prediction adaptation, the bandwidth used in a real environment would vary between what is used by each of these mechanisms alone, depending on available bandwidth and viewer head movement.

Figure~\ref{fig:eval:bandwidth_diving} presents the results for the \emph{dive} video which is chosen as it is more challenging  to predict user head movements compared to the \emph{rollercoaster} video.
Observing the results, we see that both adaptation mechanisms result in significantly less transferred data in comparison to the naive approach.
Further, we see that the popularity trace results in fair tile qualities, as the bandwidth used to download it always lies within the range of all \emph{Prediction} samples.

\begin{figure}[b]
	\includegraphics[width=0.9\linewidth]{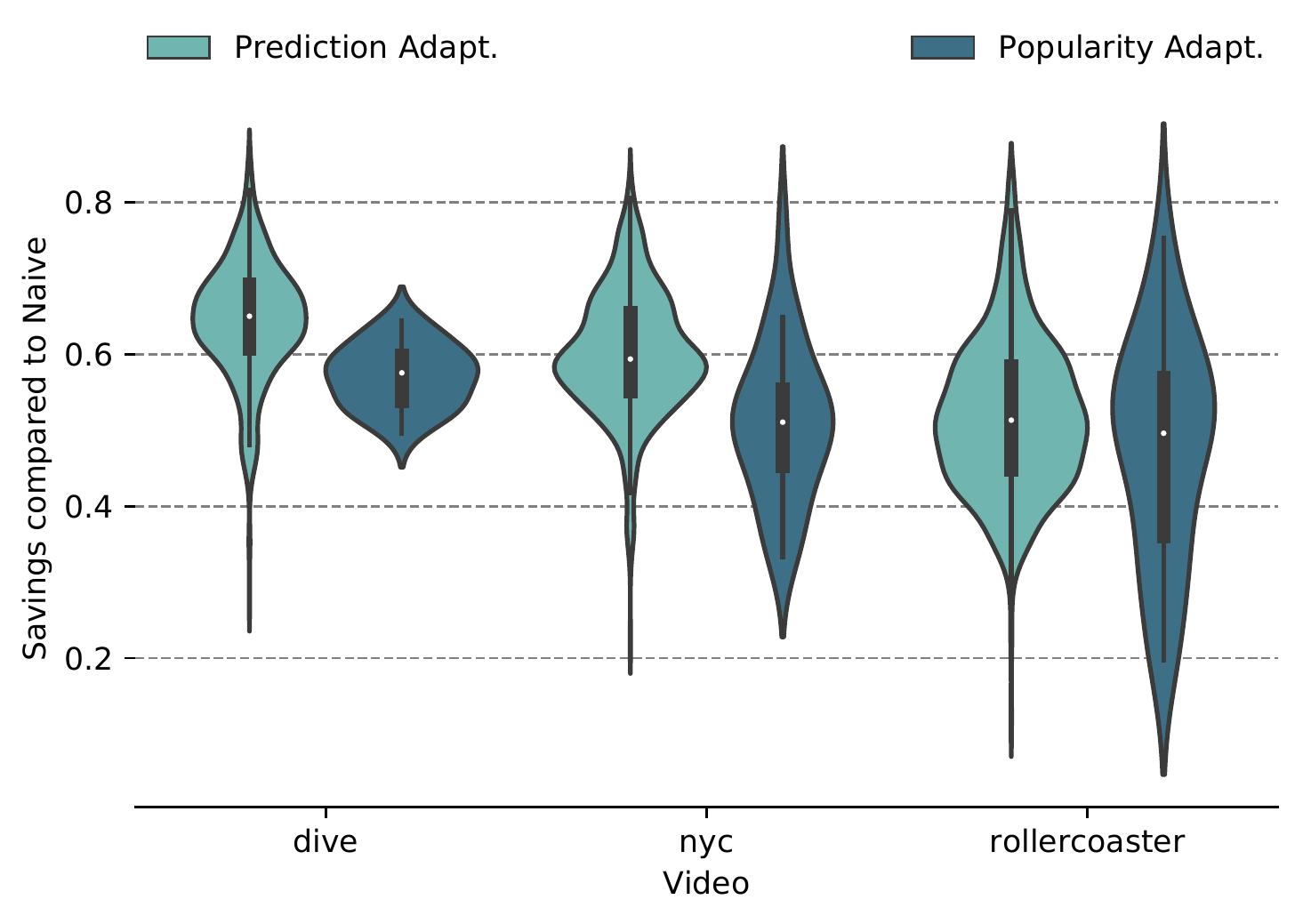}
	\caption[]{Distribution of segment-wise bandwidth savings compared to naive streaming approach.}
	\label{fig:eval:bandwidth_savings}
\end{figure}

Figure~\ref{fig:eval:bandwidth_savings} shows the distribution of segment-wise bandwidth savings of all videos, represented as violin plots.
From the width of the plots, which represent frequency, we can tell that at the majority of times, transferred data is reduced by 40\% to 70\%, depending on the video.



\begin{figure}[b]
	\centering
	\captionsetup[subfigure]{width=8cm,labelformat=empty,skip=50pt}
	\subfloat[a)]{\includegraphics[width=.5\linewidth]{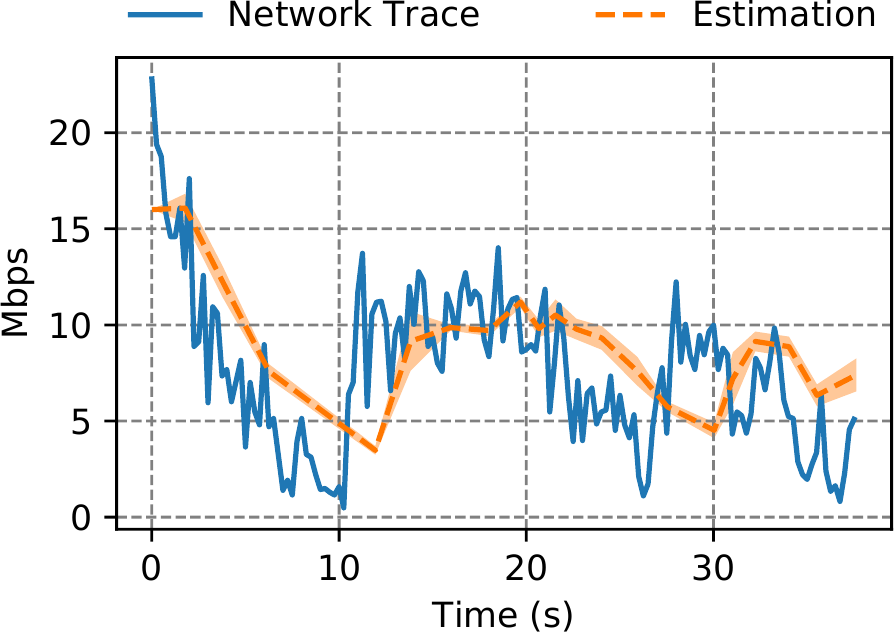}}
	\subfloat{\includegraphics[width=.5\linewidth]{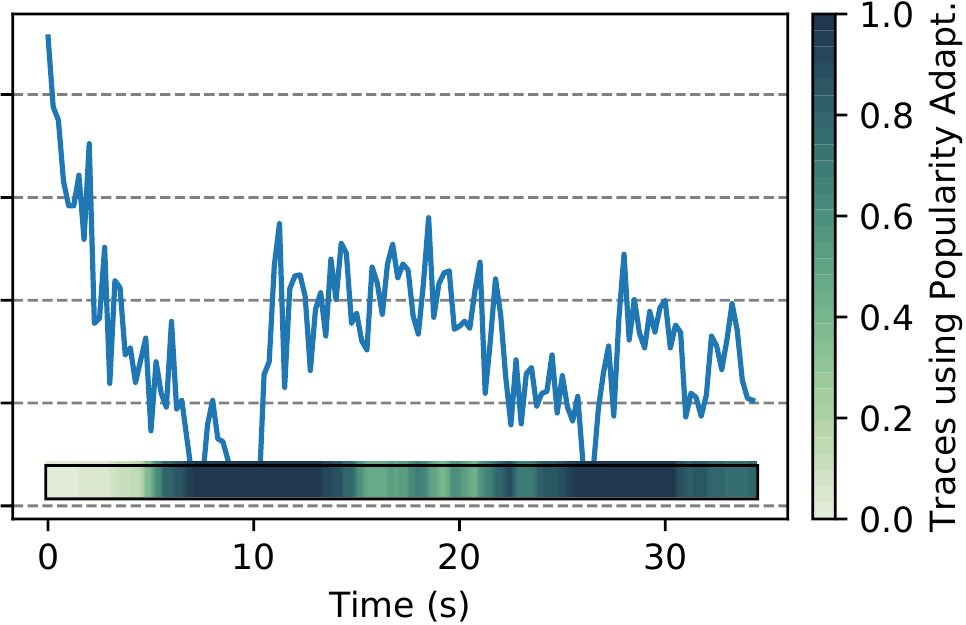}}
	
	\subfloat[b)]{\includegraphics[width=.5\linewidth]{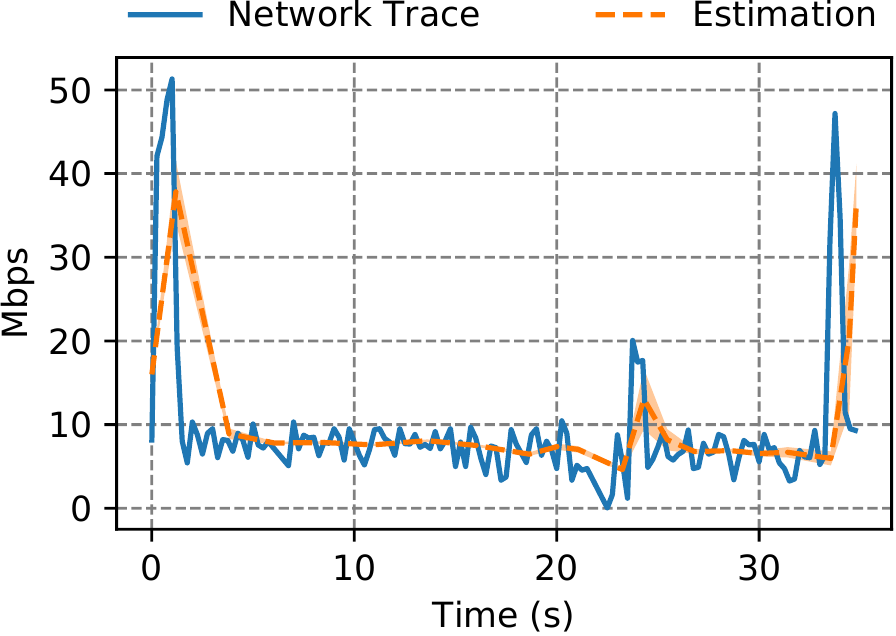}}
	\subfloat{\includegraphics[width=.5\linewidth]{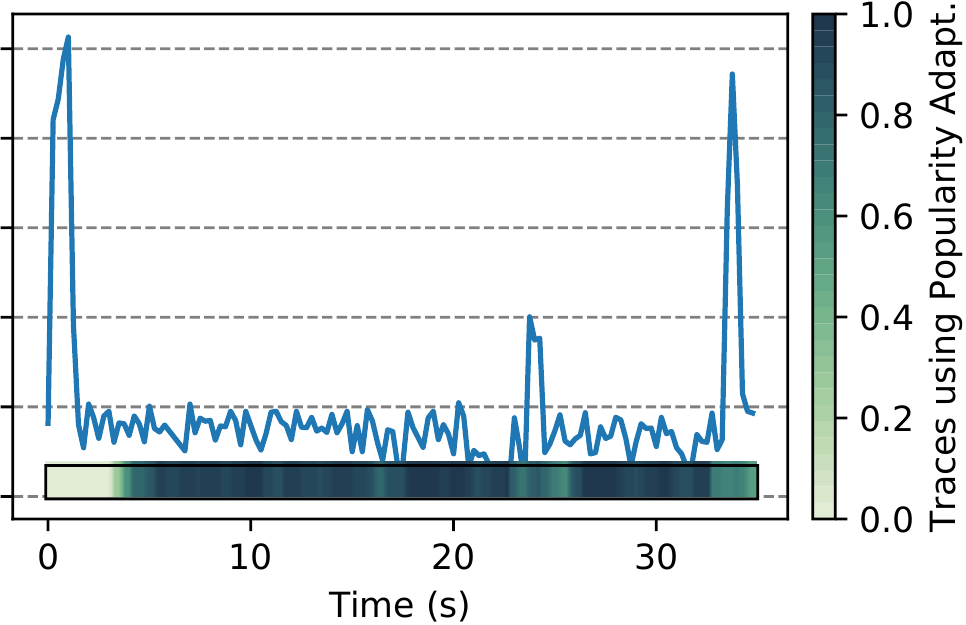}}
	
	\subfloat[c)]{\includegraphics[width=.5\linewidth]{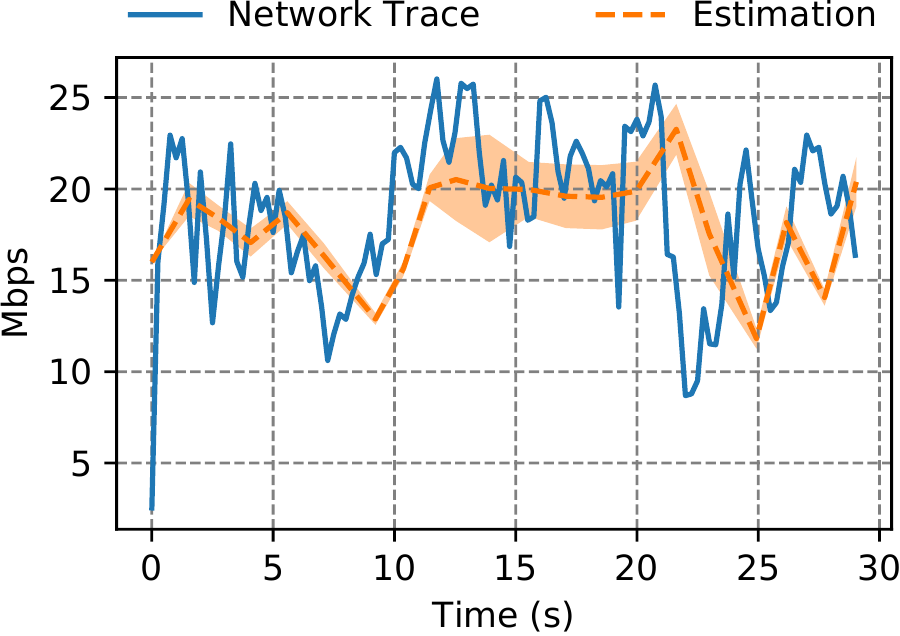}}
	\subfloat{\includegraphics[width=.5\linewidth]{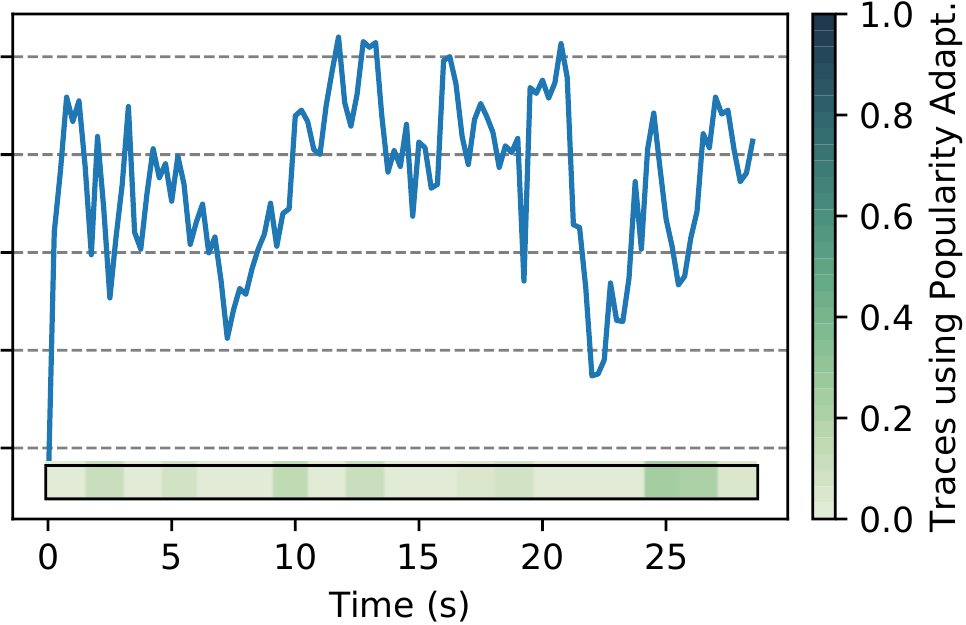}}
	
	\caption{Left: Network trace and corresponding client estimation. Right: Share of viewers using popularity adaptation, while the rest uses prediction adaptation. Top: test\_trace, mid: TMobile-LTE-driving, bottom: Verizon-LTE-driving.}
	\label{fig:eval:bandwidth_estimation_test}
\end{figure}

\begin{figure*}[t]
	\includegraphics[width=.9\textwidth]{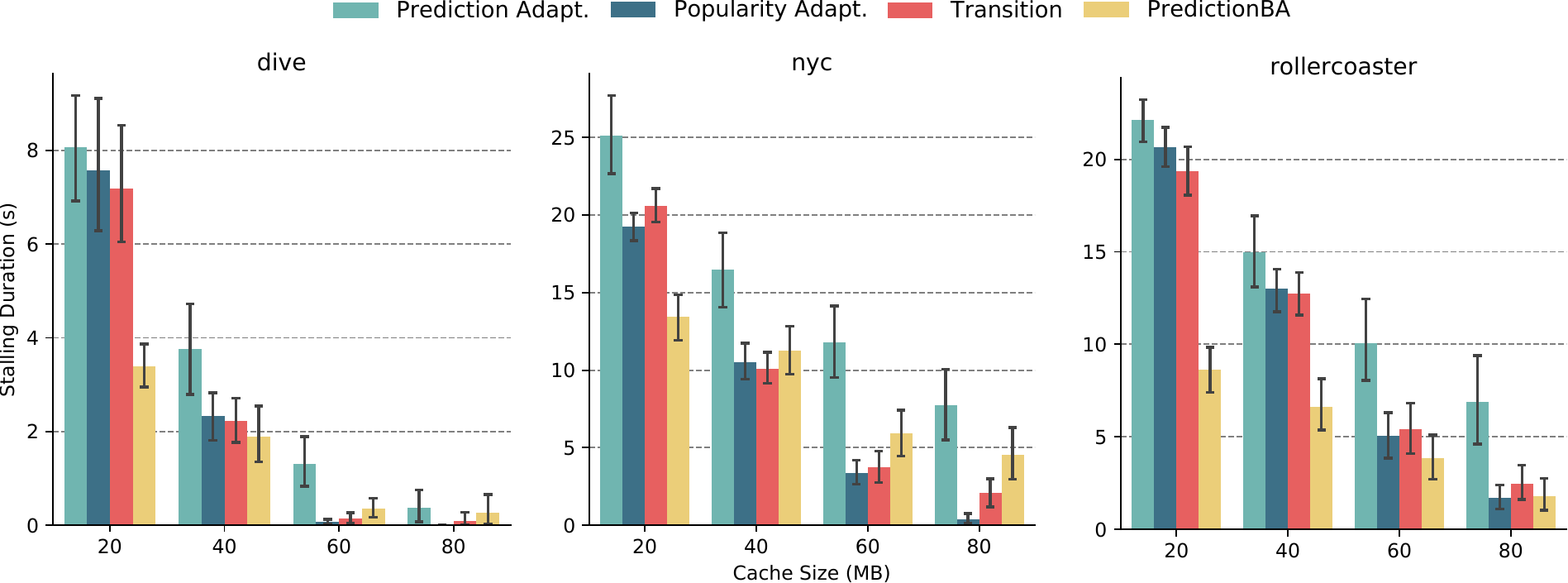}
	\caption{Distribution of stalling durations using viewers per video.}
	\label{fig:eval:stalling}
\end{figure*}

\subsection{Transitions and QoE}
Our 360° video streaming system issues a transition if the bandwidth estimate falls beyond a threshold that is determined by the currently selected tile qualities from the \emph{Prediction} adaptation policy.
Specifically, we determine the threshold to be the bandwidth required to continue streaming the currently requested video quality.
For our evaluation, a dynamic network environment is created by using recorded network traces from~\cite{winstein2013stochastic} to adapt the throughput between the server and the cache in a realistic manner.
The bandwidth between the cache, that we assume to be an edge cache, is not throttled. 
To this end we used the Linux command \emph{tc}\footnote{\url{https://linux.die.net/man/8/tc} [Accessed: \today]}.
The traces in use are \emph{Verizon-LTE-short.down}, \emph{Verizon-LTE-driving.down} and \emph{TMobile-LTE-driving.down}.
The throughputs generated from \emph{Verizon-LTE-short.down} were scaled by 1.5 to cause more transitions, which is named \emph{test\_trace}.
In Figure~\ref{fig:eval:bandwidth_estimation_test}, we show the three network traces and our system's estimate on the left, while the bandwidth traces in conjunction with the share of users using popularity adaptation is shown on the right.
%
The orange area behind the dashed estimation lines represents the standard deviation of running the simulation 30 times. As this area is generally quite small, we can assume that the server bandwidth throttling and the client's estimation algorithms operate consistently. As the client computes a bandwidth estimate only when a new segment is about to be downloaded from the source, the estimation graphs are far coarser than the actual network traces.
This also leads to the estimate seemingly lagging behind the real trace.
This can be countered by estimating the bandwidth more frequently~\cite{squad12}, e.g., between every downloaded tile.
On the right side, we empirically indicate the probability of a transition to popularity adaptation over the duration of the trace by a bar at the x-axis.
Here, a darker color indicates a higher transition probability.
These results comprise the measurements of 30 different viewing traces downloading the \emph{dive} video, while the server throttled its throughput according to the network trace used.
If the same viewing trace was used in all iterations, the results visible in the mentioned plots would be binary, i.e., either popularity or prediction adaptation would be used in all iterations at a time.
However, with different viewing traces, the bandwidth threshold responsible for triggering transitions varies depending on the predicted tile qualities.

\subsubsection{Stalling Events}
To evaluate the QoE of our streaming system, we first examine how effectively stalling events are prevented.
This is measured by the total duration of stalling events occurring during the playback of a video.
For the data acquisition, the server throttles its throughput according to the \emph{test\_trace}, as out of the three network traces, this one causes the most transitions (cf. Figure~\ref{fig:eval:bandwidth_estimation_test} a).
We repeated the experiment 30 times per video.

Additionally, in each iteration four different adaptation mechanisms were used to download the videos.
\emph{PredictionBA} uses the bitrate for \textbf{B}andwidth \textbf{A}daptation instead of transitions to mitigate throughput drops.
This acts as a benchmark for our transition scheme, as this is the traditional strategy of dealing with dynamic network conditions.
As our scheme transitions between prediction and popularity adaptation, these mechanisms were also executed and measured separately to compare their individual stalling behavior.
The results are presented in Figure~\ref{fig:eval:stalling}.
We observe that all adaptation mechanisms benefit from an increased cache size since this increases the chance of a requested file being stored on the cache.
As determined in Section~\ref{eval:bandwidth}, popularity traces tend to have higher bandwidth requirements than prediction adaptation.
Nevertheless, popularity adaptation consistently achieves better results in stalling duration than prediction adaptation, as a higher BHR is achieved.
Further, we can observe that the distribution of \emph{Transition} stalling durations always lies between the corresponding distributions of \emph{Popularity} and \emph{Prediction}.
This is also an expected observation, since with transitions we always use one of both mechanisms.
In case of stable network conditions, a \emph{Transition} distribution will converge towards either \emph{Popularity} or \emph{Prediction}.
With small cache sizes, bitrate adaptation generally achieves the best results.
This is especially noticeable with the \emph{rollercoaster} video.
To achieve a comparable result with our transition logic, the cache needs to satisfy a capacity requirement that depends on the video size.
If the cache does not meet this requirement, bitrate adaptation should be preferred over transitions, as minimum stalling is the most important requirement to provide high QoE \cite{taghavinasrabadi2017adaptive}.
To lower this threshold, either a cache replacement policy that results in a cache state, which better matches the segment popularities, or information about which files are actually stored on the cache is needed.
Thereby, popularity adaptation can achieve a higher BHR with unaltered cache sizes, and thus, also reduce stalling events.
Summarizing, our \emph{transition} approach shows a stalling performance between \emph{prediction} and \emph{popularity}.

\subsubsection{Video Quality}
\begin{figure}[t]
	\includegraphics[width=\linewidth]{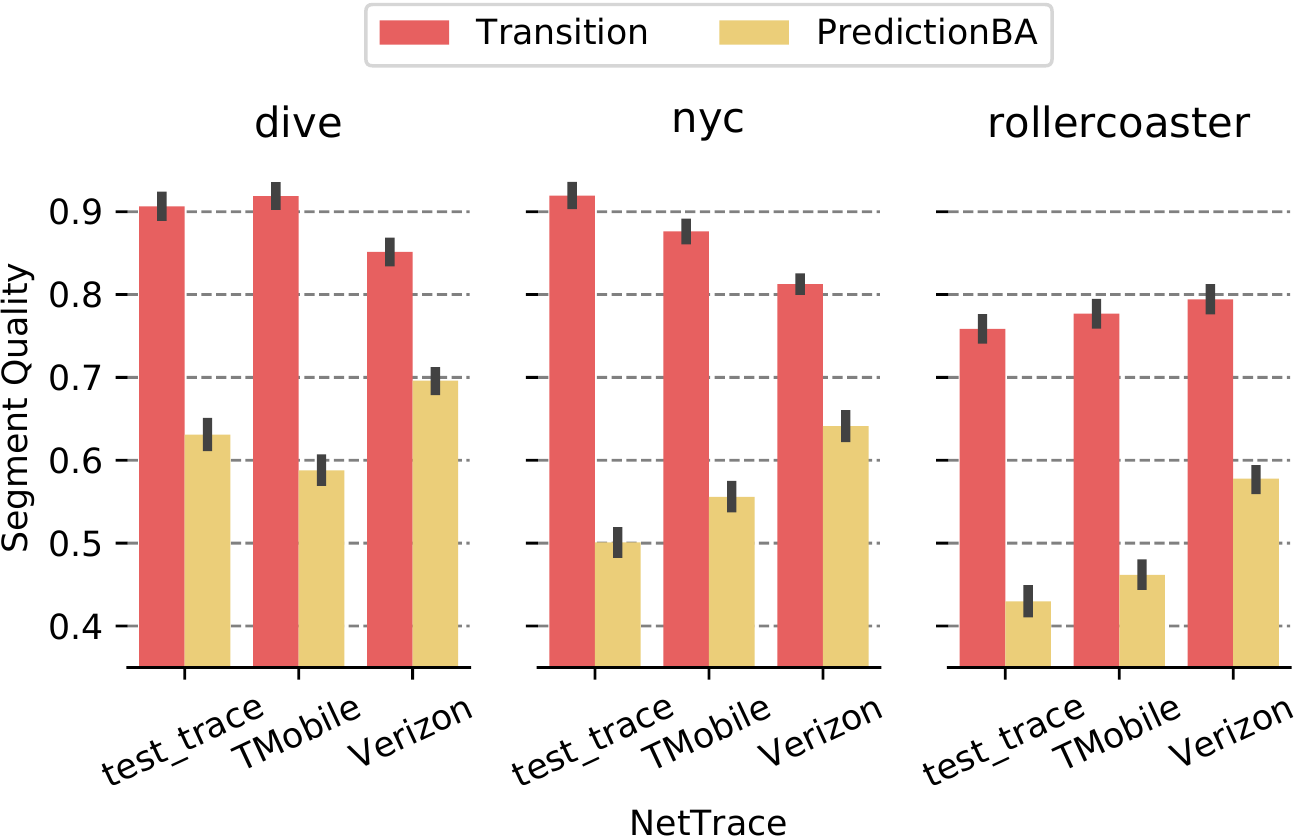}
	\caption{Average tile quality per video segment over 30 experiments per video. The y-axis denotes this quality in a range between 0 and 2, since 3 quality levels are available.}
	\label{fig:eval:quality}
\end{figure}

Next, we evaluate the effects of our scheme on video quality.
Therefore, we measure the average quality level of all tiles downloaded during an entire video.
As the videos are encoded in three quality levels, this value ranges from 0 to 2.
It is important to note that this does not represent the actual quality displayed in the HMD.
We chose not to measure this, as it would require a human reaction to popularity adaptation, i.e., a reaction to when only popular tiles are streamed in high quality, independent of head orientation.
An extensive user study would need to be conducted in order to evaluate this metric.
Nevertheless, the average overall quality does reflect the benefits our streaming scheme has over traditional bitrate adaptation.
The results are depicted in Figure~\ref{fig:eval:quality}.
We can observe that \emph{Transition} always achieves a significantly higher average quality than \emph{PredictionBA}.
Quality improvements of up to 80\% can be observed, e.g., for nyc and test\_trace.
Averaged over all videos and network traces, the quality gain amounts to 50.93\%.
The difference in quality decreases with network traces that have a higher overall throughput.
This is the case because \emph{PredictionBA} performs fewer bitrate adaptations when such network traces are in use.
Analogous to the previous evaluation, \emph{Transition} always achieves results between what is achieved by \emph{Prediction} and \emph{Popularity}.
Particularly, with the high throughput network trace \emph{Verizon-LTE-driving}, \emph{Transition} converges against \emph{Prediction}, as close to no transitions occur with these network conditions.
%

\begin{figure}[t]
	\includegraphics[width=\linewidth]{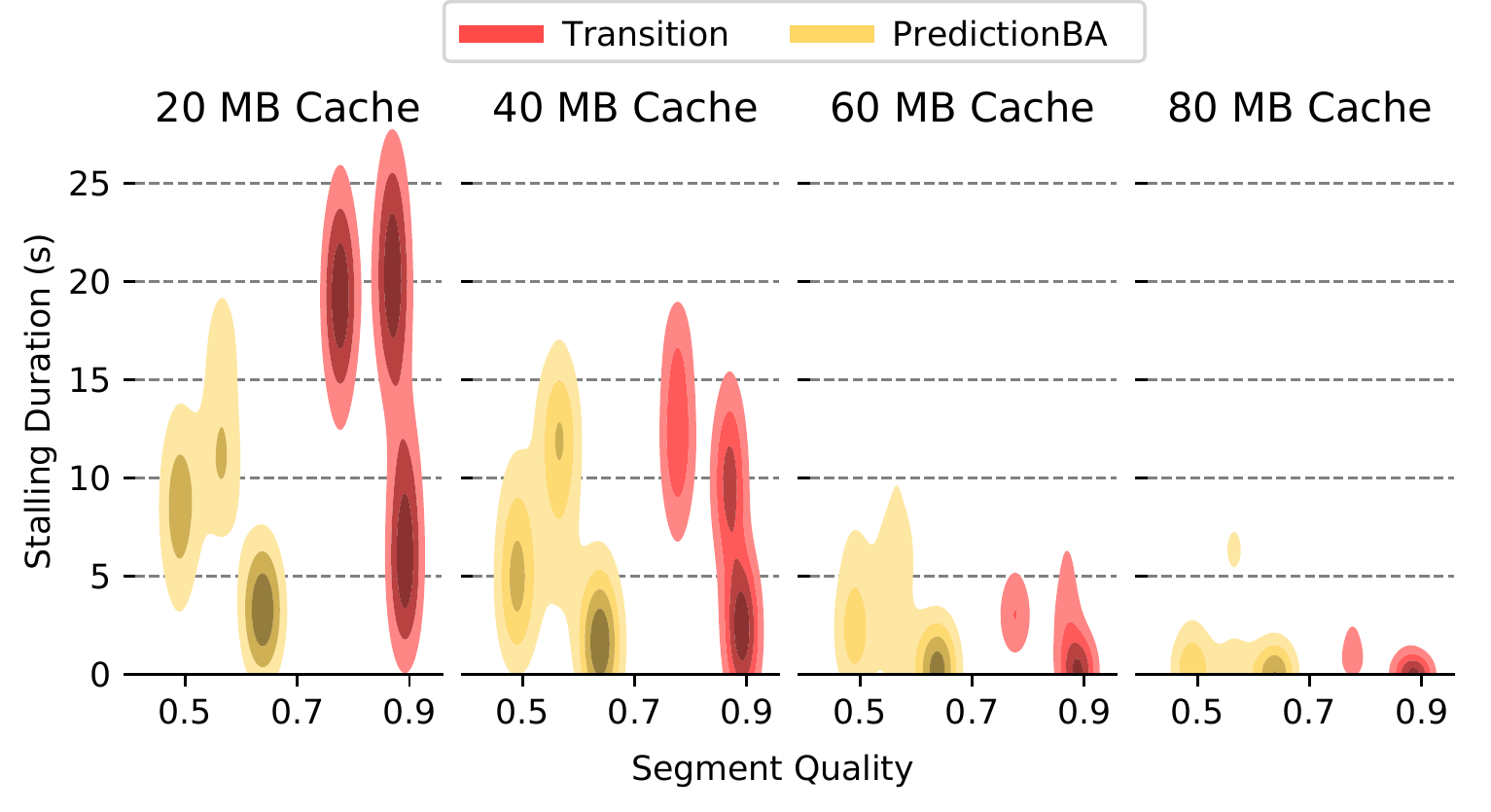}
	\caption{Distribution of stalling durations vs. segment qualities of \emph{Transition} and \emph{PredictionBA} for
		\emph{test\_trace} and all videos displayed in order (rollercoaster,nyc,dive). Darker shades represent higher density.}
	\label{fig:eval:qs}
\end{figure}

To conclude this section, we combine the previous measurements of video quality and, additionally, the staling duration over the video playback for \emph{Transition} and \emph{PredictionBA}.
Both of these aspects have a major impact on QoE, and thus, Figure~\ref{fig:eval:qs} illustrates how user-friendly both strategies perform.
To this end, we increased the cache size starting from 20 MB to 80 MB.
Note that  80 MB amounts to about 50\% of \textit{rollercoaster}'s file size.
At a cache sizes larger than 40 MB, we observe a similar stalling distribution between \textit{Transition} and \textit{PredictionBA}.
\textit{PredictionBA} achieves particularly good results for a small cache size of 20 MB, since many predicted tiles overlap with the popularity trace, as this video has a distinct focus point.
With larger sizes, \textit{Transition} outperforms \emph{PredictionBA}, as only few predicted tiles match the popularity trace.
This applies especially for dynamic content such as the \emph{nyc} video.

The three videos lead to the three visible density cores, from left to right: rollercoaster, nyc, and dive.
As minimum stalling durations and high video quality are desired, an ideal distribution would be located at the bottom right corner in each plot.
The \emph{Transition} mechanism leads to higher stalling durations than \emph{PredictionBA} with small cache sizes.
This drawback disappears with growing cache capacities, while video quality is always improved significantly independent of cache size.

\section{Discussion}\label{sec:discussion}
In the following, we discuss some limitations of our system as well as ways to mitigate these limitations:

\emph{i)} Our system does not consider that arbitrary \emph{sensory-predicted} tiles can also be stored on the intermediary cache system.
Therefore, transitions might occur too soon, as predicted tiles can be still fetched sufficiently fast, despite the slow server link.
To mitigate this, predicted tiles can be matched with the popularity trace.
Assuming these matching tiles being stored on the intermediary cache, these tiles can be considered separately when computing the required bandwidth.

\emph{ii)}
Stalling events can occur when not enough data is fetched directly from the cache whilst popularity adaptation is active.
This can happen when the intermediary cache does not dedicate enough capacity to store frequently requested content.
This issue can hardly be avoided, since we do not know which tiles are stored at  the cache.
Popularity-based adaptation only gives us an indication about which tiles are \emph{likely} to be stored at the cache.

\emph{iii)}  The bandwidth requirements provided by the manifest file do not refer to individual tile segments, but to entire tiled video streams.
This results in inaccurate estimates for the next segment's bitrate, as segments of a video differ in file size.
A solution to this problem would be to add the bitrate of each tile segment to the manifest file.

\emph{iv} To access the Quality of Experience (QoE) of the users, a user study is desirable but out of scope for this paper. A key research question for such a study is how the \emph{guided view} impacts the users' QoE.

Despite these weaknesses, our transition system improves the performance compared with bitrate adaptation, given that the cache's capacity equals or exceeds a minimum size.

\section{Conclusion and Future Work}\label{sec:conclusion}
This paper presents a 360° video streaming system that is able to transition between video quality adaptation strategies depending on the available bandwidth.
To this end, we consider a cache being placed between the content origin and the client, e.g., by a CDN or an edge cache operated by the Internet provider.
We demonstrate that the system improves the average video quality by 50.93\% in comparison to a bitrate adaptive scheme.
In scenarios with low network throughput and appropriate caching, up to 80\% higher video quality was measured.
We showed in our experiments that the presented system achieves the highest byte hit rates with the cache eviction policy LFUDA - Least Frequently Used with Dynamic Aging.
Independent of cache size and network conditions, our system consistently offers significantly higher video quality.
Additionally, our system reduces the bandwidth consumption between 40\% and 70\% in comparison to basic streaming approaches.
In future work, we plan to extend our system by making the client aware of the cache's long-living hot contents~\cite{bhat2018sabr} and assess the benefit of proactive caching~\cite{ckdiss18} for our system.
Additionally, we will address how so provision the cache in oder to minimize stalling events given a specific average video target quality.
Further, we plan to incorporate multiple popularity traces per video to use them to classify users considering the regions in the videos which they are most interested in.

\begin{acks}
	This work has been funded by the German Research Foundation (DFG) as part of the projects C3 and B4 in the Collaborative Research Center (SFB) 1053 MAKI.
\end{acks}

\balance
\bibliographystyle{ACM-Reference-Format}
\bibliography{bibliography}

\end{document}